\documentstyle[12pt]{article}
\begin{document}
\def\a{\alpha}\def\b{\beta}\def\g{\gamma}\def\d{\delta}\def\e{\epsilon }
\def\k{\kappa}\def\l{\lambda}\def\L{\Lambda}\def\s{\sigma}\def\S{\Sigma}
\def\Th{\Theta}\def\th{\theta}\def\om{\omega}\def\Om{\Omega}\def\G{\Gamma}
\def\y{\vartheta}\def\m{\mu}\def\n{\nu}
\def\ws{worldsheet}
\def\susy{supersymmetry}
\def\ts{target superspace}
\def\ks{$\k$--symmetry}
\newcommand{\plabel}{\label}
\renewcommand\baselinestretch{1.3}
\newcommand{\nn}{\nonumber\\}\newcommand{\p}[1]{(\ref{#1})}
\renewcommand{\thefootnote}{\fnsymbol{footnote}}
\thispagestyle{empty}
\begin{flushright}
{\bf FTUV-00-1002}\\
{\bf hep--th/0008249}
\end{flushright}

\vspace{1.5cm}

\begin{center}
{\Large SUPEREMBEDDING APPROACH and S-DUALITY.}

\bigskip

{\Large  A unified description of superstring and super-D1-brane.}

\vspace{2.0cm}

{\bf Igor Bandos}

\vspace{1.0cm}
{\small\it 
Institute for Theoretical Physics\\
NSC Kharkov Institute of Physics and Technology\\
Akademicheskaya 1, 61108, Kharkov,  Ukraine}\\
{\small\it 
and 
\\ 
Departamento de Fisica Teorica, \\
Facultad de Fisica, \\
Universidad de Valencia, \\ 
E-46100 Burjassot (Valencia), Spain}
\\ 
e-mail: 
{\it bandos@ific.uv.es, bandos@hep.itp.tuwien.ac.at }

\vspace{2.0cm}

{\bf Abstract}
\end{center}

{It is proved that a  basic superembedding equation for the 2-dimensional
worldsheet superspace $\S^{(2|8+8)}$ embedded into  D=10 type IIB
superspace ${\underline{\cal M}}^{(10|16+16)}$ provides  a universal,
S-duality invariant description of a
fundamental superstring and  super-D1-brane.
We work out generalized action principle,
obtain superfield equations of motion for both these objects and
find how the S-duality transformations relate the superfield
equations of superstring and super-D1-brane.

The  superembedding of 6-dimensional worldsheet
superspace $\S^{(6|16)}$ into the D=10 type IIB superspace
${\underline{\cal M}}^{(10|16+16)}$
will probably provide a similar universal description
for the set of
type IIB super--NS5--brane, super-D5-brane and a Kaluza-Klein monopole
(super-KK5-brane).}

\vspace{0.3cm}

\vspace{0.3cm}
{\bf PACs: 11.25-w, 11.30Pb, 04.65+e}

\renewcommand{\thefootnote}{\arabic{footnote}}
\setcounter{page}0
\setcounter{footnote}0

\newpage

\setcounter{equation}{0}
\def\theequation{\thesection.\arabic{equation}}

\section{Introduction}

It is believed that type IIB superstring theory has
$SL(2,Z)$ S-duality symmetry
\cite{HT,W}.
$SL(2,R)$ symmetry had been found in the low energy limit
provided by type IIB $D=10$ supergravity theory \cite{Schwarz83,HW}.
When nonperturbative BPS states including super-Dp-branes with
$p=1,3,5,7,9$ are considered, the
 Dirac charge quantization 
condition for the brane charges shall be taken into account.
It reduces $SL(2,R)$ down to $SL(2,Z)$ which, thus, appears as a
quantum S-duality group.

On the classical level, however, one can consider the continuous
group $SL(2,R)$. When superspace is flat this symmetry reduces
down to $SO(2)$ acting on Grassmannian coordinates of the flat
superspace.

Among the BPS states are ones
which are invariant under S-duality.
A well known  example is super--D3--brane \cite{SchD,KO}. The
analysis of the supergravity solution \cite{Hull}
and of the action  in the second order approximation \cite{Lozano}
indicates that Kaluza--Klein monopole 5-brane (super--KK5--brane)
should be invariant as well.  In distinction, the fundamental string and
super-D1-brane as well as the type IIB NS5-brane and super-D5-brane are
not invariant under $SL(2,Z)$ group.  The reason is that fundamental
string is coupled minimally to the NS--NS (Neveu-Schwarz--Neveu-Schwarz)
2-form gauge field $B_2$ of type IIB supergravity, while super-D1-brane is
coupled to the RR (Ramond-Ramond) gauge field $C_2$.  The type IIB NS5
superbrane and super-D5-brane carry unit magnetic charges with respect
to $B_2$ and $C_2$ fields.  The classical S-duality symmetry $SL(2,R)$
mixes $B_2$ and $C_2$ fields as well as RR and NS-NS charges.

The minimal irreducible multiplets of BPS states under the
S-duality group $SL(2,Z)$ are provided by families of so--called
$(p,q)$ strings ($(1,0)$ corresponds to the fundamental superstring and
$(0,1)$ to the super-D1-brane) \cite{Schwarz,Witten} and  $(p,q)$
5-branes \cite{LRCGNN}.
An intensive search for an S-duality invariant  universal
description of
$SL(2,Z)$ multiplets of BPS states
can be witnessed today
\cite{HK,Townsend,CT,CWW}.

The Hamiltonian analysis and some related studies
\cite{KO,HK}
indicate that the $(p,q)$ string
can be associated with D1-brane action
considered on the surface of Born--Infeld equation.
Thus, a unified description of $(p,q)$--string is basically a {\sl
universal description of fundamental superstring (super--NS1--brane) and
Dirichlet superstring (super--D1--brane)}.

The main message of this  article is that
 such a  universal, S-duality invariant description
 of the sets of superbranes
 is  provided by the superembedding approach
\cite{stv,vz,stvz,Berk,gs92,DGHS,gs93,PT},
\cite{bpstv,bsv,hs96,bst,HSopen,Dima}.
In this framework a $10$-dimensional type IIB super-(D)p-brane
is described by a
{\sl worldsheet superspace} $\Sigma^{(p+1|16)}$ embedded into the
target type IIB superspace ${\underline{\cal M}}^{(10|32)}$.
The number of fermionic 'directions' of the worldsheet superspace
($=16$)
is twice less than the one of the target superspace ($=32$).
And each of the fermionic directions is in one--to--one correspondence
\cite{stv} with a parameter of $\kappa$-symmetry
\cite{ALS} inherent to
super-p-brane actions
\footnote{Note that there exists a possibility to introduce a worldsheet
superspace with less number $n$ of the fermionic dimensions
$\Sigma^{(p+1|n)}$, $n\leq 8$.
Then the remaining $(16-n)$ $\kappa$-symmetries become nonmanifest gauge
symmetries of a superfield action, which can be built in this case
\cite{stvz,Berk,pst2,Dima}. } or, equivalently, with a parameter of
supersymmetry preserved by the BPS--state corresponding to the
super-(D)p-brane.

The embedding is specified by the so--called geometrodynamical constraint
or {\sl basic superembedding equation}.
Its  form is universal: the superembedding condition implies
that a pull--back of a bosonic supervielbein form of the
target superspace on the
worldvolume superspace has vanishing projections on the fermionic
worldvolume directions (see \cite{stv,vz,gs92,bpstv,hs96,bst,Dima} and
below).  For branes in $D=10$ type II superspaces the superembedding
equation contains dynamical equations of motion among its consequences
\cite{gs93,bpstv,bsv,hs96}.
Then the universal description of the superstring and super-D1-brane
as well as of the set of super-D5-brane, type IIB super--NS5--brane
and super-KK5-brane should occur
if i) the natural worldvolume
superspaces of these objects are of  the same type:
i.e. have the same number of bosonic and fermionic directions and the 
same worldvolume 'quantum numbers' of the fermionic coordinates,
ii) the equations of motion for all objects can be encoded
in the same superembedding condition.
The former item can be justified by studying
the structure of the $\kappa$--symmetry of the superbranes.

An investigation of the universal description  of different branes
in the frame of superembedding approach will hopefully provide us
with new insights in the structure of String/M--theory.
In this paper we begin the above program by studying the
universal, S-duality invariant description of the superstring and
super-D1-brane.

\subsection{Basic notations and short summary}

The basic equation of the superembedding approach for the type IIB
superstring specifies the embedding
\begin{equation}\plabel{Z=Z}
Z^{\underline{M}} =
\hat{Z}^{\underline{M}} (\zeta ):  \qquad
{X}^{\underline{m}} =\hat{X}^{\underline{m}} (\xi , \eta ),
~~~~{\Theta}^{\underline{\mu}1} =\hat{\Theta}^{\underline{\mu}1} (\xi ,
\eta ),
~~~~{\Theta}^{\underline{\mu}2} =\hat{\Theta}^{\underline{\mu}2} (\xi ,
\eta )
\end{equation}
of 2-dimensional worldsheet superspace
\begin{equation}\plabel{S2-16}
\Sigma^{(2|8+8)}: ~~ \{ (\zeta^M) \} =
\{ (\xi^m, \eta^{+q}, \eta^{-\dot{q}} ) \}, \quad m=0,1, ~~
~~ q=1,\ldots , 8 , ~~ \dot{q}=1,\ldots, 8
\end{equation}
into the 10-dimensional type IIB superspace
\begin{equation}\plabel{IIBss}
\underline{{\cal M}}^{(10|16+16)}= \{ (X^{\underline{m}},
\Theta^{\underline{\mu}1}, \Theta^{\underline{\mu}2}) \}, \qquad
{\underline{m}}=0,\ldots ,9 ,
\quad {\underline{\mu}}=1,\ldots ,16.
\end{equation}
It can be written in the form (see \cite{stv,gs92,DGHS,PT,bpstv,Dima} and
refs.  in \cite{Dima})
\begin{equation}\plabel{basic}
\hat{E}^{\underline{a}}_{+q} \left( \hat{Z}(\zeta) \right) = 0, \qquad
\hat{E}^{\underline{a}}_{-\dot{q}} \left( \hat{Z}(\zeta)
\right) = 0. \qquad
\end{equation}
Here
${E}^{\underline{a}}\equiv
 d{Z}^{\underline{M}}
{E}^{~\underline{a}}_{\underline{M}}
 ({Z}) $ 
is the bosonic supervielbein form of the type IIB superspace and
\begin{equation}\plabel{hEua}
\hat{E}^{\underline{a}}\equiv
 d{Z}^{\underline{M}}(\zeta )
\hat{E}^{~\underline{a}}_{\underline{M}}
 \left(\hat{Z}(\zeta)\right) =
e^{++} \hat{E}^{~\underline{a}}_{++} +
e^{--} \hat{E}^{~\underline{a}}_{--}
+ e^{+q} \hat{E}^{~\underline{a}}_{+q}
+ e^{-\dot{q}} \hat{E}^{~\underline{a}}_{-\dot{q}}
\end{equation}
is its pull-back on the worldsheet superspace  $\S^{(2|8+8)}$.
In Eq. \p{hEua}
\begin{equation}\plabel{aA}
e^A \equiv d\zeta^M e_M^{~A}(\zeta ) =
\left(e^{++}, e^{--};
e^{+q}, e^{-\dot{q}}
 \right)
\end{equation}
is an intrinsic supervielbein of the worldsheet superspace.
It can be either subject to the supergravity constraints,
or induced by the embedding \cite{bpstv,bsv}.

When the superembedding equations \p{basic} are taken into account,
the general decomposition of the pull-back of the bosonic vielbein form
\p{hEua} reduces to
\begin{equation}\plabel{se}
\hat{E}^{\underline{a}}\equiv
 d{Z}^{\underline{M}}(\zeta )
\hat{E}^{~\underline{a}}_{\underline{M}}
 \left(\hat{Z}^{\underline{M}}(\zeta)\right) =
e^{++} \hat{E}^{~\underline{a}}_{++} +
e^{--} \hat{E}^{~\underline{a}}_{--}.
\end{equation}

The superembedding equation for type IIB superstring puts the theory
on the mass shell as it does for all the objects with more then
16 target space supersymmetries, such as D=11 supermembrane  \cite{bpstv}
(super-M2-brane),  super-M5-brane  \cite{hs96}
and type IIA superstring
 \cite{gs93,bpstv}. However, as it was found in \cite{bpstv},
  the case of type IIB has a peculiarity. Namely, the equations of
  motion which follow from \p{se} contains a constant parameter $a$
  (see also \cite{gs93} for $D=3$).
  E.g., for the case of {\sl flat target type IIB superspace}
\begin{equation}\plabel{Pi}
E^{\underline{a}} = \Pi^{\underline{m}}
\d^{~\underline{a}}_{\underline{m}}, \qquad
{\Pi}^{\underline{m}} =
dX^{\underline{m}} -
i d{\Theta}^{1\underline{\mu}}
{\sigma}^{\underline{m}}_{\underline{\mu}\underline{\nu}}
{\Theta}^{1\underline{\nu}}
-
i d{\Theta}^{2\underline{\mu}}
{\sigma}^{\underline{m}}_{ \underline{\mu}\underline{\nu}}
{\Theta}^{2\underline{\nu}},
\end{equation}
\begin{equation}\plabel{EdT}
E^{\underline{\a}1} =
d{\Theta}^{1\underline{\mu}}
\d^{~\underline{\a}}_{\underline{\mu}}, \qquad
E^{\underline{\a}2} =
d{\Theta}^{2\underline{\mu}}
\d^{~\underline{\a}}_{\underline{\mu}} \qquad
\end{equation}
 the fermionic equations
  which follow from \p{se} can be written in the form
\begin{equation}\plabel{feqma}
\left(\nabla_{--} \hat{\Theta}^{\underline{\mu}1}
+ a \nabla_{--} \hat{\Theta}^{\underline{\mu}2}\right)
\s_{\underline{a}\underline{\mu}\underline{\nu}}
\hat{E}^{\underline{a}}_{++} =0, \qquad
\left(\nabla_{++} \hat{\Theta}^{\underline{\mu}2} - a \nabla_{++}
\hat{\Theta}^{\underline{\mu}1}\right)
\s_{\underline{a}\underline{\mu}\underline{\nu}}  \hat{E}^{\underline{a}}_{--}
=0, \qquad
\end{equation}
while the fermionic equations for type IIB superstring are
\begin{equation}\plabel{feqm}
\nabla_{--} \hat{\Theta}^{\underline{\mu}1}
\s_{\underline{a}\underline{\mu}\underline{\nu}}  \hat{E}^{\underline{a}}_{++}
=0, \qquad
\nabla_{++} \hat{\Theta}^{\underline{\mu}2}
\s_{\underline{a}\underline{\mu}\underline{\nu}}  \hat{E}^{\underline{a}}_{--}
=0 \qquad
\end{equation}
Eqs. \p{feqm} coincide with \p{feqma} only when the constant
$a$ vanishes.

On the other hand, if one uses an evident scale invariance of
Eqs. \p{feqma}
and multiplies them by ${1\over \sqrt{1+a^2}}$, then he
arrives at the equations related to Eqs. \p{feqm} by the
$SO(2)$ transformation of the Grassmann coordinates of the flat type IIB
superspace \cite{bpstv}

 \begin{equation}\plabel{SO(2)}
\left(\matrix{ \hat{\Theta}^{\underline{\mu}1 \prime} \cr
\hat{\Theta}^{\underline{\mu}2\prime} } \right)=
\left(\matrix{ {1\over \sqrt{1+a^2}} & {a\over \sqrt{1+a^2}} \cr
-{a\over \sqrt{1+a^2}} & {1\over \sqrt{1+a^2}} }
\right)
\left(\matrix{ \hat{\Theta}^{\underline{\mu}1} \cr
\hat{\Theta}^{\underline{\mu}2} } \right)
\equiv
\left(\matrix{ Cos~ \a & Sin~ \a \cr
- Sin~ \a & Cos~ \a \cr }
\right)
\left(\matrix{ \hat{\Theta}^{\underline{\mu}1} \cr
\hat{\Theta}^{\underline{\mu}2} } \right)
\end{equation}

The $SO(2)$ transformations \p{SO(2)} are {not} a
symmetry of the action of type IIB superstring.
They are the {\sl flat superspace image of the
S-duality group} $SL(2,R)$.
(We will make it apparent in Section 6).
Nevertheless, $SO(2)$ {\sl is} an evident symmetry of the
superembedding equation \p{basic}, \p{se}.

Thus the solutions of the superembedding equation \p{feqma} with $a\not=0$,
which are characterized by a superfield generalization of Eqs. \p{feqm}
(see below), shall describe extended objects which are related to
fundamental superstring by S-duality.  As we will demonstrate in this
paper, Eqs. \p{feqma} with $a\not=0$ are just the fermionic equations of
motion for a super-D1-brane.  The numerical parameter $a$ is expressed
through the on-shell value of the gauge field strength of the
super-D1-brane
$dA-B_2 = 1/2 e^{++} \wedge e^{--} F^{(0)}$
\begin{equation}\plabel{a(F)}
a = \pm \sqrt{{1+ F^{(0)}\over 1- F^{(0)}}},
\qquad dF^{(0)}=0.
\end{equation}
The 'scale' $\sqrt{1+a^2}$ of the parameter $a$ is
unessential when the equations are considered, while the $SO(2)$ parameter
($2\alpha $)
is essentially in one--to--one correspondence with the on--shell value of
the field strength
\begin{equation}\plabel{al(F)}
Cos~2\a = -F^{(0)}, \quad
\ Cos ~\a \equiv {a \over \sqrt{1+a^2}} = \pm \sqrt{{1+ F^{(0)}\over 2 }}.
\qquad
\end{equation}

Sections 2--5 of the present paper are devoted to the proof of  the above
result.  We motivate that the worldvolume superspace of a super-D1-brane
is of the same type as the one of the  fundamental superstring \p{S2-16}
and that the superembedding condition is the same as well \p{basic}.  We
obtain the explicit relation \p{a(F)} between the parameter $a$ and
on-shell value of the  generalized field strength of the worldvolume gauge
field and justify the statement that the value of  the field strength,
the superembedding equation and the fermionic superfield equations
specify the description of the super-D1-brane completely.

To this end we use  the generalized action principle
\cite{bsv,bst,bpst,b1,abkz} for superstring and super-D1-brane.
It is a brane counterpart of the group manifold action for
supergravity \cite{rheo}.  The advantage of the generalized action
\cite{bsv} is that it produces the
superembedding equations, (a superfield generalization  of the) proper
equations of motion and the worldvolume supergravity constraints in a
universal manner.  Moreover, it possesses a generalized
$\kappa$--symmetry \cite{b1,bst,abkz} and, hence, provides a bridge between
the standard (Green-Schwarz---Dirac-Born-Infeld) formulation
and the superembedding 
approach \cite{bpstv,hs96,Dima}.  The $\kappa$--symmetry appears in the
irreducible form and, thus, can be used to determine the properties of the
worldsheet superspace necessary for the superembedding description of the
superbranes.

\subsection{Contents of the paper}

For simplicity, in Sections 2,3,4,5 we restrict ourself by the case of flat
target superspace, where the classical S-duality  group $SL(2,R)$ reduces
to $SO(2)$.
Section 2 is devoted to the general solution of the superembedding
equation. Lorentz harmonics (moving frame variables) are introduced here.
In Section 3 and 4 we elaborate the generalized action principle and
obtain superfield equations of motion for the superstring and
super--D1--brane. The universal description of the superstring and
super--D1--brane in the frame of the superembedding approach
is considered in Section 5.
 The results for the general supergravity background are presented
in Section 6. It is demonstrated that the $SO(2)$ rotations
which, together with super--Weyl
transformations,
relate super--D1--brane and superstring, are indeed the
compensated rotations which appear when the classical S--duality group
$SL(2,R)$ acts on the unimodular matrix constructed from the axion and
dilaton superfields. We also use the super--Weyl transformations
to derive the expression for $(p,q)$--string tension
\cite{Schwarz}.  Some useful formulae are collected in the Appendices.

\setcounter{equation}{0}

\section{General solution of superembedding equation. The role of
                   Lorentz harmonics.}

\subsection{Lorentz harmonics and superembedding equation}

To obtain in a regular manner all the consequences
of the superembedding equation \p{basic} it is convenient to use its
equivalent form \cite{bpstv} (see Appendix A)

\begin{equation}\plabel{seEI}
\hat{E}^{I}\equiv
\hat{E}^{\underline{b}} U^{~I}_{\underline{b}}=0, \qquad I=1,\ldots 8,
\end{equation}
where $U^{~I}_{\underline{a}}$ are $8$ orthogonal and normalized
10-vector superfields
\begin{equation}\plabel{UIUJ}
U^{\underline{a}I}
U_{\underline{a}}^J = - \d^{IJ}.
\end{equation}
Their  set can be completed up to a basis of $10$-dimensional
(tangent) space, or, equivalently, to the Lorentz group
valued matrix ({\sl moving frame matrix})
\begin{equation}\plabel{U1}
U^{(\underline{b})}_{\underline{a}}
=
\left( U^{0}_{\underline{a}},
U^{~J}_{\underline{a}},
U^{9}_{\underline{a}}\right)
~ \in ~ SO(1,9) \quad \Leftrightarrow \quad
U^{(\underline{b})}_{\underline{a}}
U^{\underline{a}(\underline{c})} = \eta^{(\underline{b})(\underline{c})}
\equiv \left(\matrix{ 1 & 0         & 0  \cr
		      0 & - \d^{IJ} & 0 \cr
		      0 & 0         & -1\cr}\right)
\end{equation}
 by adding 2 orthogonal and normalized vectors
$U^{0}_{\underline{a}}$,
$U^{9}_{\underline{a}}$
\begin{equation}\plabel{U0U9} U^{\underline{a}0}
U_{\underline{a}}^0 = 1, \quad
U^{\underline{a}0}
U_{\underline{a}}^9 = 0,
\quad
U^{\underline{a}9}
U_{\underline{a}}^9 = -1. \qquad
U^{\underline{a}0}
U_{\underline{a}}^I = 0 =
U^{\underline{a}9}
U_{\underline{a}}^I.
\end{equation}

It is convenient to replace
the vectors
$U^{0}_{\underline{a}}$,
$U^{9}_{\underline{a}}$
by their light--like combinations
\begin{equation}\plabel{U+-}
U^{++}_{\underline{a}}=  U^{0}_{\underline{a}}+ U^{9}_{\underline{a}},
\qquad
 U^{--}_{\underline{a}}=  U^{0}_{\underline{a}}-
U^{9}_{\underline{a}}
\end{equation}
$$
U^{++}_{\underline{a}} U^{++\underline{a}}=0, \qquad
U^{--}_{\underline{a}} U^{--\underline{a}}=0, \qquad
U^{++}_{\underline{a}} U^{--\underline{a}}=2.
$$
and to define the moving frame matrix \p{U1} by
\begin{equation}\plabel{vhII}
U^{(\underline{b})}_{\underline{a}}
=
\left(
U^{++}_{\underline{a}},
U^{--}_{\underline{a}},
U^{~J}_{\underline{a}}
\right)
~ \in ~ SO(1,9) \quad \Leftrightarrow \quad
U^{(\underline{b})}_{\underline{a}}
U^{\underline{a}(\underline{c})} = \eta^{(\underline{b})(\underline{c})}
\equiv \left(\matrix{ 0 & 2  & 0  \cr
		      2 & 0  & 0  \cr
		      0 & 0  &  - \d^{IJ} \cr}
		      \right)
\end{equation}
Its elements can be recognized as {\sl Lorentz harmonics},
introduced by Sokatchev \cite{Sok}.

The Lorentz rotation by matrix \p{vhII} provides us with the bosonic
supervielbein forms
\begin{equation}\plabel{E(a)}
\hat{E}^{(\underline{a}^\prime)}\equiv
\hat{E}^{\underline{b}}
U^{~(\underline{a}^\prime)}_{\underline{b}}
= \left(\hat{E}^{++}, \hat{E}^{--}, \hat{E}^{I}\right),
\qquad
\hat{E}^{\pm\pm}=
\hat{E}^{\underline{a}}
U^{\pm\pm}_{\underline{b}},
\quad
\hat{E}^{I}=
\hat{E}^{\underline{a}}
U^{I}_{\underline{b}}
\end{equation}
adapted to the superembedding in the sense of Eq. \p{seEI}.
To adapt the complete supervielbein (more precisely, the pull-back of the
supervielbein) to the superembedding, we need in the double covering
of the Lorentz group valued matrix $U$ \p{vhII}. It is given by
$16\times 16$ matrix
\begin{equation}\plabel{shII}
V^{(\underline{\alpha}^\prime) }_{\underline{\a}}
= \left(\matrix{
V^{~+}_{\underline{\a}q}\cr 
V^{~-}_{\underline{\a}\dot{q}}\cr }\right)
\quad \in \quad Spin(1,9)
\end{equation}
whose rectangular $16\times 8$ blocks
$
V^{+{q}}_{\underline{\a}},
V^{-\dot{q}}_{\underline{\a}}
$ ($q=1,\ldots 8,$ $\dot{q}= 1, \ldots 8$) are called
{\sl spinor Lorentz harmonics} \cite{B90,gds,BZ,bpstv,bsv} or
spinor moving frame variables \cite{BZ}.
The claimed statement that \p{shII} provides a double covering of the
Lorentz rotation described by \p{vhII}
can be expressed by the conditions of $\gamma$-matrix preservation
\footnote{The constraints \p{vss} are certainly reducible. However, they
are most convenient for calculations. The irreducible form of the constraints
for spinor harmonics can be found in \cite{gds,BZ}.}
\begin{equation}\plabel{vss}
 U^{(\underline{a})}_{\underline{b}}
\sigma^{\underline{b}}_{\underline{\a}\underline{\b}}
= V_{\underline{\a}}^{~(\underline{\g})}
\sigma^{(\underline{a})}_{(\underline{\g})(\underline{\d})}
V^{(\underline{\d})}_{\underline{\b}}, \qquad
\end{equation}
With an appropriate $SO(1,1) \otimes SO(8)$ invariant representation
for $D=10$ $\s$-matrices, the constraints \p{vss} can be split into the
set of the following covariant relations
 \begin{equation}\plabel{vssA1}
 U^{++}_{\underline{a}}
\sigma^{\underline{a}}_{\underline{\a}\underline{\b}}
= 2V_{\underline{\a}q}^{~+}
V^{~+}_{\underline{\b}q}, \qquad
 U^{++}_{\underline{a}}
\tilde{\sigma}^{\underline{a}\underline{\a}\underline{\b}}
= 2V^{+\underline{\a}}_{\dot{q}}
V_{\dot{q}}^{+\underline{\b}},  \qquad
\end{equation}
\begin{equation}\plabel{vssA2}
 U^{--}_{\underline{a}}
\sigma^{\underline{a}}_{\underline{\a}\underline{\b}}
= 2V_{\underline{\a}\dot{q}}^{~-}
V^{~-}_{\underline{\b}\dot{q}}, \qquad
 U^{--}_{\underline{a}}
\tilde{\sigma}^{\underline{a}\underline{\g}\underline{\b}}
= 2V^{-\underline{\g}}_{q}
V_{q}^{-\underline{\b}}, \qquad
\end{equation}
\begin{equation}\plabel{vssA3}
 U^{I}_{\underline{a}}
\sigma^{\underline{a}}_{\underline{\a}\underline{\b}}
= 2V_{(\underline{\a}{q}}^{~+} \g^I_{q\dot{q}}
V^{~-}_{\underline{\b})\dot{q}}, \qquad
 U^{I}_{\underline{a}}
\tilde{\sigma}^{\underline{a}\underline{\g}\underline{\b}}
= - 2V^{-(\underline{\g}}_{{q}} \g^I_{q\dot{q}}
V_{\dot{q}}^{+\underline{\b})},  \qquad
\end{equation}
which imply, in particular, that the spinor harmonics
$
V^{+{q}}_{\underline{\a}},
V^{-\dot{q}}_{\underline{\a}}$
can be treated as square roots from the light--like vectors
$U^{++}_{\underline{a}},
U^{--}_{\underline{a}}$.

The second equalities in Eqs. \p{vssA1}, \p{vssA2}, \p{vssA3}
involve the inverse Lorentz harmonics \cite{BZ}
\begin{equation}\plabel{shII-1}
V^{-\underline{\a}}_{p}
V^{~+}_{\underline{\a}q}= \delta_{pq}, \qquad
V^{-\underline{\a}}_{p}
V^{~-}_{\underline{\a}\dot{q}} =0,
\end{equation}
$$
V^{+\underline{\a}}_{\dot{p}}
V^{~+}_{\underline{\a}q}=0, \qquad
V^{+\underline{\a}}_{\dot{p}}
V^{~-}_{\underline{\a}\dot{q}}
=\delta_{\dot{p}\dot{q}}.
$$

\medskip

The supervielbein adapted for the superembedding in the sense of Eq.
\p{seEI} is
\begin{equation}\plabel{EAIIB}
 (\hat{E}^{(\underline{a}^\prime)};~~ \hat{E}^{(\underline{\a}^\prime) 1},
~~\hat{E}^{(\underline{\a}^\prime ) 2})=
\left(\hat{E}^{++},\hat{E}^{--},
\hat{E}^{I}; ~
\hat{E}^{+{q} 1}, \hat{E}^{-\dot{q} 1}, ~
\hat{E}^{+{q} 2}, \hat{E}^{-\dot{q} 2}\right),
\end{equation}
where
\begin{equation}\plabel{E+q}
\hat{E}^{+q 1, 2} =
\hat{E}^{\underline{\a}1,2} V^{~+}_{\underline{\a}q},
\qquad
\hat{E}^{-\dot{q} 1, 2} =
\hat{E}^{\underline{\a}1,2}
V^{~-}_{\underline{\a}\dot{q}}.
\end{equation}
For the case of flat type IIB
superspace Eqs. \p{E+q} become
\begin{equation}\plabel{E+q0}
\hat{E}^{+q 1, 2} =
d\hat{\Theta}^{\underline{\mu}1,2} V^{~+}_{\underline{\mu}q},
\qquad
\hat{E}^{-\dot{q} 1, 2} = d\hat{\Theta}^{\underline{\mu}1,2}
V^{~-}_{\underline{\mu}\dot{q}}.
\end{equation}

The geometry of the worldsheet superspace can be specified by embedding
in the following sense \cite{bsv}.
One can choose the bosonic components of the worldsheet supervielbein
\p{aA} to
be equal to the $E^{\pm\pm}$ components of the adapted supervielbein
\p{EAIIB} (see Appendix A)
\begin{equation}\plabel{E=e}
\hat{E}^{++}\equiv
\hat{E}^{\underline{b}} U^{++}_{\underline{b}}=e^{++}, \qquad
\hat{E}^{--}\equiv
\hat{E}^{\underline{b}} U^{--}_{\underline{b}}=e^{--}. \qquad
\end{equation}
The fermionic worldsheet supervielbein forms can be identified
with 16 linear combinations of the pull-backs of target
superspace fermionic forms \p{E+q}. We make the choice
\begin{equation}\plabel{eq=Eq}
e^{+q} = E^{+q1} =
\hat{E}^{\underline{\a}1} V^{+q}_{\underline{\a}},
\qquad
e^{-\dot{q}}= \hat{E}^{-\dot{q}2} =
\hat{E}^{\underline{\a}2}
V^{-\dot{q}}_{\underline{\a}},
\end{equation}
which will be motivated below by a study of the irreducible $\k$--symmetry.
In the case of {\sl flat type IIB superspace} the worldsheet spin
connections and $SO(8)$ gauge field can be identified with Cartan forms
\cite{BZ,bpstv,bsv}
\begin{equation}\plabel{omAIJ}
 \om
  \equiv {1 \over 2}
U^{--}_{\underline{a}} d U^{++\underline{a}}, \qquad
 A^{IJ}
  \equiv
U^{I}_{\underline{a}} d U^{J\underline{a}}
\end{equation}
while the remaining Cartan forms
\begin{equation}\plabel{fpmpmI}
f^{\pm\pm I} \equiv
U^{\pm\pm}_{\underline{a}} d U^{I\underline{a}}
\end{equation}
are covariant with respect to local $SO(1,1) \times SO(8)$
transformations
\footnote{In general supergravity background  one can define the
worldsheet connections and covariant forms
by adding to the expressions \p{omAIJ}, \p{fpmpmI}
the supergravity spin connections 
$w_{\underline{a}}^{~\underline{b}}$ contracted with the harmonic vectors
$$
\tilde{f}^{\pm\pm I} \equiv f^{\pm\pm I} +
(UwU)^{\pm\pm I} \equiv
U^{\pm\pm}_{\underline{a}}
(d U^{I\underline{a}} + w^{~\underline{b}}_{\underline{a}}
U_{\underline{b}}^{I}), \qquad
 \tilde{\om}
  \equiv  \om + {1 \over 2} (UwU)^{-- | ++}, \qquad
 \tilde{A}^{IJ} = {A}^{IJ} + (UwU)^{IJ}
$$
Actually, this is a prescription for the construction of the
'gauge fields of nonlinear realization' \cite{CWZ}. }.
This $SO(1,1)\times SO(8)$ is an evident symmetry of the
relations \p{seEI}, \p{E=e}, \p{eq=Eq} which acts on the bosonic
worldsheet supervielbein and vector harmonics as follows
\begin{equation}\plabel{SO(1,1)U}
e^{\pm\pm \prime} = e^{\pm\pm } exp{(\pm 2\b (\zeta )) }, \qquad {} \qquad 
U^{\pm\pm\prime}_{\underline{a}}=
U^{\pm\pm}_{\underline{a}} exp{(\pm 2\b (\zeta )) }, \qquad
\end{equation}
$$
U^{I}_{\underline{a}}\rightarrow U^{I\prime}_{\underline{a}}=
U^{J}_{\underline{a}} {\cal O}^{JI}, \qquad
{\cal O}^{JK}{\cal O}^{IK} = \d^{IJ}.
$$
Such symmetry makes possible to
consider the vectors $U^{\pm\pm}_{\underline{a}}, U^{I}_{\underline{a}}$
(constrained by \p{vhII}) as homogeneous coordinates of the
coset
 \begin{equation}\plabel{coset}
{SO(1,9)\over SO(1,1) \otimes SO(8)} =
\left\{
(U^{\pm\pm}_{\underline{a}},
U^{I}_{\underline{a}}) \right\}
\end{equation}
and to identify them with Lorentz harmonic variables \cite{Sok}.

The derivatives of the harmonic variables, which do not break
the constraints \p{vhII} are expressed through the Cartan forms
\begin{equation}\plabel{DU++0}
{\cal D}U^{++}_{\underline{a}}\equiv
 dU^{++}_{\underline{a}}- U^{++}_{\underline{a}} {\om }
=
U^{I}_{\underline{a}} {f}^{++I}, \qquad
{\cal D}U^{--}_{\underline{a}}\equiv
 dU^{--}_{\underline{a}}+ U^{--}_{\underline{a}} {\om } =
U^{I}_{\underline{a}} {f}^{--I},
\end{equation}
\begin{equation}\plabel{DUI0}
{\cal D}U^{I}_{\underline{a}}\equiv
 dU^{I}_{\underline{a}}+ U^{J}_{\underline{a}} {A}^{JI} =
{1 \over 2} U^{--}_{\underline{a}} {f}^{++I}
+ {1 \over 2} U^{++}_{\underline{a}} {f}^{--I}.
\end{equation}

\subsection{General solution of superembedding equation
in flat type IIB superspace.}

To investigate the consequences of the superembedding equation
\p{seEI} one can study its integrability conditions \cite{bpstv}
\begin{equation}\plabel{IC0}
\hat{E}^I=0 \qquad \Rightarrow \quad
d\hat{E}^I \equiv d \hat{E}^{\underline{a}} U^I_{\underline{a}}(\zeta) +
\hat{E}^{\underline{a}} \wedge dU^I_{\underline{a}}(\zeta) =  0.
\end{equation}
The integrability conditions for 'conventional constraints' \p{E=e},
\p{eq=Eq} should determine the worldsheet torsion forms.

In the flat superspace Eq. \p{IC0} acquires the form
\begin{equation}\plabel{DEI=0}
{\cal D}\hat{E}^I
\equiv d \hat{E}^{I} + \hat{E}^{J} \wedge {A}^{IJ} =
\hat{T}^{\underline{a}} U^{~I}_{\underline{a}}
+{ 1\over 2} \hat{E}^{++} \wedge {f}^{--I} +
{ 1\over 2} \hat{E}^{--} \wedge {f}^{++I} =
\end{equation}
 $$
= -i \left(e^{+{q}} \wedge \hat{E}^{-\dot{q}1} - i e^{-\dot{q}} \wedge
\hat{E}^{+q2} \right) \g^I_{q \dot{q}}+
{ 1\over 2} \hat{E}^{++} \wedge {f}^{--I} +
{ 1\over 2} \hat{E}^{--} \wedge {f}^{++I} = 0.
$$
Note that the same equation occurs in curved type IIB superspace with the
standard choice of torsion constraints
$$ T^{\underline{a}} \equiv
{\cal D}E^{\underline{a}} \equiv dE^{\underline{a}}- E^
{\underline{b}}\wedge w_{\underline{b}}^{~\underline{a}} =
-i
(E^{\underline{\a}1} \wedge E^{\underline{\b}1} +
E^{\underline{\a}2} \wedge E^{\underline{\b}2})
\s^{\underline{a}}_{\underline{\a}\underline{\b}},
$$
but with covariantized Cartan forms defined in Appendix B
($f^{\pm\pm I} \rightarrow f^{\pm\pm I} +
U^{\pm\pm}wU^I$,
see also footnote 3).

 Substituting
the most general expression for the pull-backs of the
fermionic supervielbein forms  $\hat{E}^{-\dot{q}1}$ and $\hat{E}^{+{q}2}$
\begin{equation}\plabel{E-q1}
\hat{E}^{-\dot{q}1}= e^{+p} \chi_{p\dot{q}}^{--} +
e^{-\dot{p}} h_{\dot{p}}^{~\dot{q}} +
e^{\pm\pm} \Psi_{\pm\pm \dot{q}} ^{~~~-},
\qquad
\hat{E}^{+{q}2}= e^{+p} h_{p}^{~q} +
e^{-\dot{p}}
\chi_{\dot{p}q}^{++}
+ e^{\pm\pm} \Psi_{\pm\pm {q}} ^{~~~+}
\end{equation}
 into \p{DEI=0}, one
finds the expression for $f^{\pm\pm I}$ and  the
restrictions on the coefficient superfields
$\chi, h, \Psi$.
Then the integrability conditions for
Eqs. \p{E-q1} as well as the Maurer Cartan equations
\begin{equation}\plabel{Df+}
{\cal D}{f}^{++I}\equiv  d{f}^{++I} - {f}^{++I}\wedge
{\om} + {f}^{++J} \wedge A^{IJ} = 0,
\end{equation}
\begin{equation}\plabel{Df-}
{\cal D} {f}^{--I}\equiv  d{f}^{--I} + {f}^{--I}\wedge
{\om} + {f}^{--J} \wedge A^{IJ} = 0,
\end{equation}
\begin{equation}\plabel{dom}
d\tilde{\om}= {1 \over 2} {f}^{--I}\wedge {f}^{++I},
\end{equation}
\begin{equation}\plabel{dA}
F^{IJ} = d{A}^{IJ} + {A}^{IK} \wedge {A}^{KJ} = {f}^{--[I}\wedge
{f}^{++J]}
\end{equation}
shall be investigated \footnote{
Eqs. \p{Df+}, \p{Df-}, \p{dom}, \p{dA} appear as integrability conditions
for Eqs.  \p{fpmpmI}, \p{omAIJ}. Note that they give rise to a superfield
generalization of the Peterson--Codazzi, Gauss and Rici equations of the
surface theory.
In curved superspace Eqs. \p{Df+}- \p{dA} acquire the form (see Appendix B)
$$ {\cal D}\tilde{f}^{\pm\pm I}=   U^{\underline{a}\pm\pm}
R_{\underline{a}}^{~\underline{b}} U^{I}_{\underline{b}}, \qquad
d\tilde{\om}= {1 \over 2}
\tilde{f}^{--I}\wedge \tilde{f}^{++I} +
 {1 \over 2} U^{\underline{a}--} R_{\underline{a}}^{~\underline{b}}
U^{++}_{\underline{b}}, \quad
\tilde{F}^{IJ} =
\tilde{f}^{--[I}\wedge \tilde{f}^{++J]} +
 U^{\underline{a}I} R_{\underline{a}}^{~\underline{b}}
U^{J}_{\underline{b}}.
$$}.
The calculations are simplified essentially due to the theorems about
dependence of components of Eqs. \p{DEI=0}, \p{E-q1},
\p{Df+}--\p{dA} (see Appendix in Ref. \cite{bpstv}
and refs. therein).

The complete investigation of the equations
\p{DEI=0}, \p{E-q1},
\p{Df+}--\p{dA} for the case of flat target $D=10$ type $IIB$ superspace
has been performed in \cite{bpstv}.
For our consideration it is essential that, in the frame of
the above mentioned assumptions \p{eq=Eq}, \p{omAIJ}, the general solution
of the superembedding equation for the fermionic forms is
\begin{equation}\plabel{E-q1gs}
\hat{E}^{-\dot{q}1} - a \hat{E}^{-\dot{q}2} =
 \hat{E}^{++} \Psi_{++ \dot{q}} ^{~~-},
\qquad
\hat{E}^{+{q}2}
+ a \hat{E}^{+{q}1} =
 \hat{E}^{--} \Psi_{-- {q}} ^{~~+},
\end{equation}
where $a$ is a constant parameter $da=0$.
In other words, the following fermionic equations are
contained in the list of consequences of the superembedding equation
\p{seEI} 
\begin{equation}\plabel{dTh1a}
\hat{E}^{++} \wedge
(\hat{E}^{-\dot{q}1} - a \hat{E}^{-\dot{q}2}) =0, \qquad
\hat{E}^{--} \wedge
(\hat{E}^{+2{q}}
+ a \hat{E}^{+1{q}}) =0
\end{equation}

The bosonic equations of motion are specified completely by the
fermionic equations and the superembedding condition \cite{bpstv}.

To clarify the meaning of the numerical parameter $a$ in the general
solution \p{E-q1gs} we
are going to use the generalized action principle  and obtain
 superfield equations of motion for the superstring and super-D1-brane.

\bigskip

\setcounter{equation}{0}

\section{
Generalized action and superfield equations
of motion
for type IIB superstring}

\subsection{Generalized action}

The generalized action for D=10 type IIB superstring is
defined as an integral of
a Lagrangian form ${\hat{\cal L}}_2 $ over a bosonic surface
${\cal M}^2$ embedded into the worldsheet superspace

\begin{equation}\plabel{SIIB}
S_{IIB}=  \int_{{\cal
M}^{2}} {\hat{\cal L}}_2
= \int_{{\cal M}^{2}} \left( {1 \over 2}
e^{{1\over 2} \Phi (\hat{Z})} \hat{E}^{++} \wedge  \hat{E}^{--} - \hat{B}_2 \right).
\end{equation}

\begin{equation}\plabel{M2}
{\cal M}^{2} \in \S^{(2|8+8)}: \qquad
\eta^{+q} = \eta^{+q}(\xi^m), \quad
\eta^{-\dot{q}} = \eta^{-\dot{q}}(\xi^m). \quad
\end{equation}
All the field variables entering the Lagrangian form ${\hat{\cal L}}_2 $
should be considered as superfields, but taken on the bosonic surface
\p{M2}. E.g. the embedding of ${\cal M}^2$ into the
target superspace is determined by
\begin{equation}\plabel{ZM2}
{\cal M}^2 \in \underline{{\cal M}}^{(10|16+16)}: \quad
Z^{\underline{M}} =
\hat{Z}^{\underline{M}} (\zeta (\xi) ) =
\hat{Z}^{\underline{M}} (\xi, \eta (\xi) ) \quad
\Leftrightarrow \quad \cases{
{X}^{\underline{m}} =\hat{X}^{\underline{m}}
(\xi , \eta(\xi) ), \cr {\Theta}^{\underline{\mu}1}
=\hat{\Theta}^{\underline{\mu}1} (\xi , \eta(\xi) ), \cr
{\Theta}^{\underline{\mu}2} =\hat{\Theta}^{\underline{\mu}2} (\xi ,
\eta (\xi)).}
\end{equation}
The fermionic fields $\eta^{+q}(\xi^m), \eta^{-\dot{q}}(\xi^m)$ can be
regarded as worldsheet Goldstone fermions \cite{VA}.
As their  kinetic term is absent  in \p{SIIB},
the worldsheet supersymmetry is not broken.
However, their presence is important, because
it provides the possibility
to treat the equations of motion derived from the action \p{SIIB}
as {\sl superfield equations} \cite{bsv} (see also below).

 The Lagrangian form $\hat{{\cal L}}_2$ of type IIB superstring \p{SIIB}
is constructed from the pull--backs of the supervielbein
forms of type IIB superspace \p{Pi}, \p{EdT} and
light--like vector Lorentz harmonics
$
 U^{++}_{\underline{m}}(\zeta ),
 U^{--}_{\underline{m}}(\zeta )
$ \p{vhII}.
The first term is essentially the weight product
of two bosonic supervielbeins from the adapted frame \p{E(a)} while
the second term contains
the pull--back of NS-NS superform $B_2$
($\hat{B}_2=
{1 \over 2}
d\hat{Z}^{\underline{M}}(\zeta )
\wedge
dZ^{\underline{N}} (\zeta )
B_{\underline{N}{\underline{M}}}(\hat{Z}(\zeta ) )$)
with the flat superspace value
\begin{equation}\plabel{B2}
 B_2 = i \Pi^{\underline{m}}
 \wedge \left(
d{\Theta}^{1}
{\sigma}_{\underline{m}}
{\Theta}^{1} -
i d{\Theta}^2
{\sigma}_{\underline{m}}
{\Theta}^2\right)
 +
d{\Theta}^{1}
{\sigma}^{\underline{m}}
{\Theta}^{1} \wedge
d{\Theta}^2
{\sigma}_{\underline{m}}
{\Theta}^2.
\end{equation}
$\hat{\Phi} = \Phi (\hat{Z}(\zeta))$ is
the image of the
dilaton superfield $\Phi (Z)$ on the worldsheet superspace.
It vanishes in the flat type IIB superspace.

Thus, in flat superspace \p{Pi}, \p{EdT}, \p{B2}
the Lagrangian form of the generalized action \p{SIIB} is
  \begin{equation}\plabel{L2}
\hat{{\cal L}}_2 =
 {1 \over 2} \hat{E}^{++} \wedge \hat{E}^{--} -
 i \hat{\Pi}^{\underline{m}}
 \wedge \left(
d\hat{\Theta}^{1}
{\sigma}_{\underline{m}}
\hat{\Theta}^{1} -
i d\hat{\Theta}^2
\tilde{\sigma}_{\underline{m}}
\hat{\Theta}^2\right)
 +
d\hat{\Theta}^{1}
{\sigma}^{\underline{m}}
\hat{\Theta}^{1} \wedge
d\hat{\Theta}^2
\tilde{\sigma}_{\underline{m}}
\hat{\Theta}^2.
\end{equation}

\subsection{Variation of generalized action.}

The variations of the generalized action
\p{SIIB}, \p{L2} can be derived easily with the use
of the seminal formula for the Lee derivative
  \begin{equation}\plabel{Lee}
\d {{\cal L}}_2 =
i_\d d{{\cal L}}_2 +
d (i_\d {{\cal L}}_2)
\end{equation}
where the symbol $i_\d$ is defined by
$i_\delta d{Z}^{\underline{M}}\equiv \delta \hat{Z}^{\underline{M}}$.
The derivatives of the harmonic variables shall be regarded as ones
\p{DU++0}--\p{DUI0} which do not break the constraints \p{vhII}.
Thus
$$i_\d dU^{++} \equiv
\d U^{++}=
U^I i_\delta f^{++I}
+ U^{++} i_\d \om \; ,
\qquad
i_\d dU^{--}\equiv  \d U^{--}
= U^I i_\delta f^{--I}
+ U^{--} i_\d \om \; , $$
$$i_\d dU^{I} \equiv
\d U^{I}=
{1 \over 2}
U^{--} i_\delta f^{++I}
+
{1 \over 2}
U^{++} i_\delta f^{--I} - U^J i_\d A^{JI}\; .
$$
The contractions of the Cartan forms $i_\d \om, i_\d A^{IJ},
i_\d f^{\pm\pm I}$ shall be considered as independent variations of the
harmonic variables.

  It is convenient to replace the basis
of 'holonomic variations'
$\d \hat{Z}^{\underline{M}}$ by
 \begin{equation}\label{vB}
 i_\d E^{\pm\pm} = \d \hat{Z}^{\underline{M}}
E_{\underline{M}}^{\pm\pm} (\hat{Z}), \qquad
i_\d E^{I} = \d \hat{Z}^{\underline{M}}
E_{\underline{M}}^{I} (\hat{Z}),
\end{equation}
$$
i_\d E^{+q1,2} = \d \hat{Z}^{\underline{M}}
E_{\underline{M}}^{+q1,2} (\hat{Z}), \qquad
i_\d E^{-\dot{q}1,2} = \d \hat{Z}^{\underline{M}}
E_{\underline{M}}^{-\dot{q}1,2} (\hat{Z}).$$
In the flat superspace \p{Pi}, \p{EdT} the basis \p{vB} becomes
 \begin{equation}\label{vB0}
i_\d E^{\pm\pm} \equiv
i_\d \Pi^{\underline{m}} U^{\pm\pm}_{\underline{m}}, \qquad
i_\d E^I \equiv
i_\d \Pi^{\underline{m}} U^{I}_{\underline{m}}, \qquad
\end{equation}
$$
i_\d E^{-\dot{q} 1} =
\d {\Theta}^{\underline{\mu}1}
V^{-\dot{q}}_{\underline{\mu}}, \qquad
i_\d E^{+q 2} =
\d{\Theta}^{\underline{\mu}2} V^{+q}_{\underline{\mu}},
$$
where
$$
i_\d \Pi^{\underline{m}} \equiv
\d X^{\underline{m}} -
i \d {\Theta}^{1\underline{\mu}}
{\sigma}^{\underline{m}}_{\underline{\mu}\underline{\nu}}
{\Theta}^{1\underline{\nu}}
-
i \d {\Theta}^{2\underline{\mu}}
{\sigma}^{\underline{m}}_{ \underline{\mu}\underline{\nu}}
{\Theta}^{2\underline{\nu}}.
$$

\medskip

\subsection{Superfield equations of motion and gauge symmetries.}

With the above notations \p{EAIIB}, \p{E+q}, \p{fpmpmI}
the equations of motion which follow from
the generalized action \p{SIIB} --\p{ZM2} in the flat target superspace
are

\begin{equation}\plabel{EI}
\hat{E}^{I} \equiv
  \hat{\Pi}^{\underline{m}} U^{I}_{\underline{m}}=0, \qquad
\end{equation}
\begin{equation}\plabel{dTh1}
\hat{E}^{++} \wedge  \hat{E}^{-\dot{q}1} \equiv
  \hat{\Pi}^{\underline{m}} \wedge
  d\hat{\Theta}^{1\underline{\mu}}
  V^{~-}_{\underline{\mu}\dot{q}} U^{++}_{\underline{m}}=0, \qquad
\end{equation}
\begin{equation}\plabel{dTh2}
\hat{E}^{--} \wedge  \hat{E}^{+2{q}} \equiv
  \hat{\Pi}^{\underline{m}} \wedge
  d\hat{\Theta}^{2\underline{\mu}}
  V^{~+}_{\underline{\mu}q} U^{--}_{\underline{m}}=0,  \qquad
\end{equation}
\begin{equation}\plabel{MI}
\hat{M}^I_2 \equiv \hat{E}^{--} \wedge f^{++I} - \hat{E}^{++} \wedge
f^{--I} - 4i \left(\hat{ E}^{+q1} \wedge \hat{E}^{-\dot{q}1} -
\hat{E}^{+2q} \wedge \hat{E}^{-2\dot{q}} \right) \gamma^I_{q\dot{q}} = 0.
\end{equation}

The variations of the Goldstone fermions
$\eta^{+q}(\xi^m), \eta^{-\dot{q}}(\xi^m)$ do not produce independent
equations \cite{bsv}.
Thus Eqs. \p{EI}--\p{MI} are satisfied on {\sl any}
bosonic surface ${\cal M}^2$ \p{M2} in the worldsheet superspace $\S^{(2|16)}$
\p{S2-16}. As the set of all such bosonic surfaces covers the whole
superspace $\S^{(2|16)}$, one concludes that Eqs. \p{EI}- \p{MI}
can be treated as {\sl superfield equations for the differential
forms defined on the whole worldsheet superspace} \cite{bsv}.
In this case all the field variables in \p{EI}-\p{MI} shall be treated
as worldsheet superfields
$\hat{X}=\hat{X}(\zeta )=
\hat{X}(\xi, \eta),~ \hat{\Theta}^{1,2}=\hat{\Theta}^{1,2}(\zeta)=
\hat{\Theta}^{1,2}(\xi, \eta)$,
${U}= {U}(\zeta)= {U}(\xi, \eta)$ without restrictions to a bosonic
surface.  Thus the generalized action principle \cite{bsv,bst} can be used
as a dynamical basis for the superembedding approach
\cite{bpstv,bsv,hs96,Dima}.

For our study it is especially important that Eqs. \p{dTh1}, \p{dTh2}
coincide with \p{dTh1a} taken with $a=0$.

When the fermionic worldsheet
coordinates (Goldstone fields) vanish
$\eta=0$, the set of equations \p{EI}--\p{MI} becomes
equivalent to the standard equations of motion for the
type IIB Green-Schwarz superstring
\cite{BZ}.
In this case the basic fermionic forms can be  decomposed on the
bosonic ones
$$  {\cal M}_0^2: \qquad
{E}^{+{q}2} =
E^{++}{E}_{++}^{+{q}2} + E^{--}{E}_{--}^{+{q}2},
\qquad {E}^{-\dot{q}1} =
E^{++}{E}_{++}^{-\dot{q}1} + E^{--}{E}_{--}^{-\dot{q}1},
$$
and one can write Eqs.
\p{dTh1} in the form \p{feqm} with $e^{\pm\pm}=E^{\pm\pm}$.

\medskip

The external derivative of the Lagrangian form can be used for
derivation of Eqs. \p{EI}--\p{MI} and contains as well
an information about local (gauge) symmetries of the superstring.
In flat target superspace it is
\begin{equation}\plabel{dLIIB}
d\hat{{\cal L}}^{IIB}_2 =
-2i \hat{E}^{++} \wedge  \hat{E}^{-\dot{q}1} \wedge  \hat{E}^{-\dot{q}1}
+ 2i \hat{E}^{++}
\wedge  \hat{E}^{+{q}2} \wedge  \hat{E}^{+{q}2} +
{1 \over 2} \hat{E}^I \wedge \hat{M}^I_2 + \propto \hat{E}^I \wedge
\hat{E}^J, \end{equation}
where $\hat{M}_2^I$ is defined in Eq. \p{MI}.

Eqs. \p{EI}-\p{MI} appear as a result of the variations
with respect to $i_\d f^{\pm\pm I}$,
$
i_\d E^{-\dot{q} 1} =
\d {\Theta}^{\underline{\mu}1}
V^{-\dot{q}}_{\underline{\mu}}$,
$
i_\d E^{+q 2} =
\d{\Theta}^{\underline{\mu}2} V^{+q}_{\underline{\mu}}$ and
$ i_\d E^I \equiv i_\d \Pi^{\underline{m}} U^{I}_{\underline{m}}$.
The remaining (super)field variations do not produce any
nontrivial equations and, thus, can be identified with the
parameters of {\sl local (gauge) symmetries of the model}.
Evident  gauge symmetries of the action \p{SIIB} are $SO(1,1) \times SO(8)$
($i_\delta \omega , ~ i_\delta A^{IJ}$)
and reparametrization ($i_\delta E^{\pm\pm}$).
The parameters of the {\sl stringy $\kappa$--symmetry} can be identified
with the contractions of those fermionic forms which are absent
in the first line of Eq. (\ref{dLIIB})
\begin{equation}\plabel{kIIB}
\kappa^{+q}
 \equiv i_\delta \hat{E}^{+q1} =
  \delta \hat{\Theta}^{1\underline{\mu}} V^{+q}_{\underline{\mu}}, \qquad
\kappa^{-\dot{q}}
\equiv i_\delta \hat{E}^{-2\dot{q}} =
  \delta \hat{\Theta}^{2\underline{\mu}} V^{-\dot{q}}_{\underline{\mu}}.
\end{equation}
The second equalities  of Eqs. \p{kIIB} are vailed for the flat target
superspace only. In this case the $\k$--symmetry transformations of the
generalized action are defined by
\begin{equation}\plabel{k1}
\d_\k  \hat{\Theta}^{\underline{\mu}1}(\zeta(\xi))
= \k^{+q}(\xi)  V^{-\underline{\mu}}_q(\zeta(\xi)),
\qquad
\d_\k \hat{\Theta}^{\underline{\mu}2}(\zeta(\xi))=
\k^{-\dot{q}}(\xi) V_{\dot{q}}^{+\underline{\mu}}(\zeta(\xi)).
\end{equation}
$$
 i_\k \Pi^{\underline{m}} = 0 \quad \Leftrightarrow \quad
\d_\k X^{\underline{m}} =
i \d_\k {\Theta}^{1\underline{\mu}}
{\sigma}^{\underline{m}}_{\underline{\mu}\underline{\nu}}
{\Theta}^{1\underline{\nu}}
+
i \d_\k {\Theta}^{2\underline{\mu}}
{\sigma}^{\underline{m}}_{ \underline{\mu}\underline{\nu}}
{\Theta}^{2\underline{\nu}}
$$
and by the quite complicated transformations of harmonic variables
\begin{equation}\label{dKU}
\d_\k U^{\pm\pm}_{\underline{m}} = U^I_{\underline{m}} i_\k f^{\pm\pm I},
\qquad
\d_\k U^{I}_{\underline{m}} =
{1\over 2}
U^{++}_{\underline{m}} i_\k f^{--I}+
{1\over 2}
U^{--}_{\underline{m}} i_\k f^{++I}
\end{equation}
whose explicit form is unessential for our consideration\footnote{
To find
$\d_\k U^{\pm\pm}, \d_\k U^{I}$,
 one has to solve equation $i_\k (d{\cal L}_2)=0$
with respect to $i_\k f^{--I}$, $i_\k f^{++I}$ under the suppositions
\p{k1}, $i_\k \om =0, i_\k A^{IJ}=0$.}.

In accordance with \cite{stv,Dima}, the worldsheet supersymmetry has to
be in one--to--one correspondence with the $\k$--symmetry.
Thus  the parameters of the worldsheet supersymmetry $\e$ should have the
same $SO(1,1) \times SO(8)$ 'quantum numbers' as the parameters of the
irreducible $\k$--symmetry.
Moreover, on the bosonic surface ${\cal M}^2$ \p{ZM2} one should arrive at
\begin{equation}\plabel{e=k}
\e^{+q} = \k^{+q} + \ldots, \qquad
\e^{-\dot{q}} = \k^{-\dot{q}} + \ldots. \qquad
\end{equation}

On the other hand, the parameters of the local worldsheet supersymmetry
can be identified with contractions of the worldsheet fermionic
supervielbein forms. Thus the 'quantum numbers' ($SO(1,1)$ weight and
$SO(8)$ representation) of the fermionic worldsheet supervielbein are
fixed by $\k$--symmetry as it is indicated in \p{aA}
$$
\e^{+q} = i_\d e^{+q} = \d \zeta^M e_M^{+q} \qquad
\e^{-\dot{q}} = i_\d e^{-\dot{q}} = \d \zeta^M e_M^{-\dot{q}}.
\qquad $$
Moreover, the local supersymmetry transformations of the fermionic
supervielbein
\begin{equation}\plabel{ve+q}
\d e^{+q} = D i_\d e^{+q} + i_\d (De^{+q}) = D\e^{+q} + \ldots ,
\qquad
\d e^{-\dot{q}} = D i_\d e^{-\dot{q}} + i_\d (De^{-\dot{q}}) =
D\e^{-\dot{q}} + \ldots
\end{equation}
 basically coincide with the $\kappa$--symmetry transformations of the
 induced supervielbein forms \p{eq=Eq} {\sl lifted to the whole
 worldsheet superspace}
\begin{equation}\plabel{ke+q}
\d_\k e^{+q} =
\d_\k
(d\Th^{\underline{\mu}1}
V^{~+}_{\underline{\mu}q}) =
D \k^{+q} + \ldots , \qquad
\d_\k e^{-\dot{q}} =
\d_\k (d\Th^{\underline{\mu}2}
V^{~-}_{\underline{\mu}\dot{q}}) =
D\k^{-\dot{q}} + \ldots
\end{equation}
Thus the choice of proper fermionic supervielbein forms \p{aA} is
motivated by the
{\sl irreducible} $\kappa$--symmetry of the generalized action of type IIB
superstring.

The main conclusions of our consideration in the
present section are as follows:
\begin{itemize}
\item the superfield fermionic equations of motion \p{dTh1}
for type IIB superstring
coincide with the general solution of the superembedding equations \p{dTh1a}
 taken with $a=0$.
\item the parameters of stringy
$\kappa$--symmetry can be identified as in Eq. \p{kIIB}. This indicates
that a half of $\Th^1$ and a half of $\Th^2$ (but not $\Th^1$ or $\Th^2$
completely) can be gauged away by the $\kappa$-symmetry
in a component
approach. In the superfield approach one can instead fix the gauge
$\Th^{+q1}\equiv
\Th^{\underline{\mu}1}V^{~+}_{\underline{\mu}q} = \eta^{+q}$,
$\Th^{-\dot{q}2}\equiv
\Th^{\underline{\mu}2}V^{~-}_{\underline{\mu}\dot{q}} = \eta^{-\dot{q}}$.

\item The $\kappa$--symmetry determines the proper choice of the fermionic
supervielbeine forms of the worldsheet superspace. In particular the forms
induced by the embedding in accordance with Eqs. \p{eq=Eq} are appropriate.
\end{itemize}

\setcounter{equation}{0}

\section{
Super--D1--brane.
Generalized action, superfield equations and gauge symmetries}

The generalized action for super-D1-brane (Dirichlet superstring)
has the form
\begin{equation}\plabel{SD1}
S_{D1}=  \int_{{\cal
M}^{1+1}} {\hat{\cal L}}^{D1}_2 =
\int_{{\cal M}^{1+1}}  {1 \over 2}
e^{-{1\over 2} \hat{\Phi}} \hat{E}^{++} \wedge
\hat{E}^{--}\sqrt{1-(F^{(0)})^2} +
\end{equation}
$$ + \int_{{\cal M}^{1+1}} \left[Q_0 \left( e^{-{1\over 2} \Phi}
\left(dA - \hat{B}_2\right)
- {1 \over 2}  \hat{E}^{++} \wedge  \hat{E}^{--} F^{(0)} \right) +
\hat{C}_2 + \hat{C}_0 (dA - \hat{B}_2) \right],
$$
where, for simplicity,
we put tension equal to unity $T=1$ (as we did in Eq. \p{SIIB} for the
fundamental superstring). In \p{SD1} $\hat{C}_0\equiv C_0
 \left(\hat{Z}(\xi, \eta (\xi )  )\right)$ is the axion superfield
restricted to the surface ${\cal M}^2$, $\hat{C}_2$ is the pull-back of
the RR superform and $Q_0=Q_0(\xi, \eta (\xi ) )$ is a Lagrange multiplier
superfield.  The remaining notations are the same as in  Section 3.

Eq. \p{SD1} can be obtained from the general formula of Ref. \cite{bst}
by substitution $F_{ab} = \e_{ab} F^{(0)}$ for the two--dimensional
auxiliary tensor superfield.  Then the Born-Infeld square root
acquires the form $
\sqrt{-det(\eta_{ab}+ F_{ab})} =
\sqrt{1-(F^{(0)})^2}$, which indicates that
the scalar superfield $F^{(0)}$ lives in the interval $-1< F^{(0)} <1$.

Varying the generalized action \p{SD1} with respect to the
auxiliary field $F^{(0)}$ one finds the expression
\begin{equation}\plabel{Q0}
Q_0 =- e^{-{1 \over 2} \hat{\Phi}} { F^{(0)}\over  \sqrt{1-(F^{(0)})^2} },
\end{equation}
for the Lagrange multiplier
$Q_0$. The variation  $\d Q_0$ produces
the superfield constraint for the 2-dimensional
worldsheet gauge superfield
\begin{equation}\plabel{F=F}
e^{-{1 \over 2} \hat{\Phi}} \left(dA - \hat{B}_2\right)  = {1 \over 2}
\hat{E}^{++}
\wedge \hat{E}^{--} F^{(0)},
\end{equation}
while the  variation of  ({\sl a priori} unconstrained) gauge superfield
$A=d \zeta^{M}(\xi) A_M(\zeta(\xi))$ results in
the 2-dimensional Born-Infeld equation
\begin{equation}\plabel{BI}
d\left( Q_0 e^{-{1\over 2}\hat{\Phi}} + \hat{C}_0
\right)
=0.
\qquad
\end{equation}
 In flat superspace \p{Pi}, \p{EdT}, where $C_0=0$, $\Phi =0$ and
\begin{equation}\plabel{C2}
 C_2 = i \Pi^{\underline{m}}
 \wedge \left(
d{\Theta}^{2}
{\sigma}_{\underline{m}}
{\Theta}^{1} -
i {\Theta}^2
{\sigma}_{\underline{m}}
d{\Theta}^1\right)
 +
(d{\Theta}^{1}
{\sigma}^{\underline{m}}
d{\Theta}^{1}
-
d{\Theta}^{2}
{\sigma}^{\underline{m}}
d{\Theta}^{2})
\wedge
{\Theta}^2
{\sigma}_{\underline{m}}
{\Theta}^1,
\end{equation}
  Eq. \p{BI} reduces to
\begin{equation}\plabel{dQ0=0}
dQ_0= 0 \qquad \Rightarrow \qquad dF^{(0)} =0.
\end{equation}
and implies that
the field strength $F^{(0)}$ of the worldsheet gauge field is a constant.

The variations of the harmonics
($\d U^{\pm\pm} = U^I i_\d f^{\pm\pm I}$) result in the
{\it same superembedding equation as for the
case of fundamental superstring}
\begin{equation}\plabel{EID1c}
\hat{E}^{I} \equiv
  \hat{E}^{\underline{a}} U^{I}_{\underline{a}}=0. \qquad
\end{equation}

If one considers  the derivative of the
Lagrangian form  taken on the surface of
algebraic and geometric equations  \p{Q0}, \p{F=F}, \p{EID1c}
$d{{\cal L}}^{D1}_2\vert_s$, after some tedious but straightforward
algebraic manipulations one finds that it contains only 16 of 32
independent fermionic forms $\hat{E}^{2+}_{~{q}}, \hat{E}^{1+}_{~{q}},
\hat{E}^{2-}_{~\dot{q}}, \hat{E}^{1-}_{~\dot{q}}$
(see Appendix B), namely 
\begin{equation}\plabel{E2-E1}
\left( \hat{E}^{2-}_{~\dot{q}} -
{\sqrt{1-F^{(0)} \over 1+ F^{(0)}}} \hat{E}^{1-}_{~\dot{q}}\right),
\qquad
\left( \hat{E}^{2+}_{~{q}} +
{\sqrt{1+F^{(0)} \over 1 - F^{(0)}}} \hat{E}^{1+}_{~{q}}\right).
\end{equation}
This is the reflection of
the fermionic $\kappa$-symmetry on the level of Noether identities.
The parameters of the $\kappa$-symmetry can be identified
with the contractions of some linear combinations of the forms
which are independent on
\p{E2-E1}. In particular we can define
\begin{equation}\plabel{kD1}
\tilde{\kappa}^{1+{q}}= i_\d \hat{E}^{1+}_{~{q}}, \quad
\tilde{\kappa}^{2-}_{~\dot{q}} = i_\d \hat{E}^{1-}_{~\dot{q}}, \quad
\Leftrightarrow \quad
\tilde{\kappa}^{\underline{\a}} = \d \hat{Z}^{\underline{M}}
\hat{E}^{\underline{\a}1}_{\underline{M}}(\hat{Z}),
\end{equation}
In the case of flat target superspace Eq. \p{kD1} becomes
$
\tilde{\kappa}^{\underline{\mu}} = \d \hat{\Theta}^{\underline{\mu}1}
$
and means that the Grassmann coordinate field
$\hat{\Theta}^{\underline{\mu}1}$
can be gauged away by $\kappa$--symmetry
in a component approach or identified with Grassmann coordinate
of the worldsheet superspace
$
\Th^{\underline{\mu}1}=
\eta^{\underline{\mu}}$ in the superfield approach.

On the other hand, {\sl the parameter of the irreducible $\k$--symmetry
can be defined in the same way as in the case of type IIB superstring}
\p{kIIB}
$$
\kappa^{+q}
 \equiv i_\delta \hat{E}^{+q1} =
  \delta \hat{\Theta}^{1\underline{\mu}} V^{+q}_{\underline{\mu}}, \qquad
\kappa^{-\dot{q}}
\equiv i_\delta \hat{E}^{-2\dot{q}} =
  \delta \hat{\Theta}^{2\underline{\mu}} V^{-\dot{q}}_{\underline{\mu}}.
$$
Thus the worldsheet superspace with the geometry induced by embedding in
accordance with Eqs. \p{E=e}, \p{eq=Eq}, \p{omAIJ} is proper for the
description of both the superstring and super-D1-brane.

\bigskip

\setcounter{equation}{0}

\section{Universal description of superstring and super-D1-brane by
superembedding. Flat type IIB superspace. }

In flat type IIB superspace the superfield generalization of the fermionic 
equations of motion
of the super-D1-brane
\begin{equation}\plabel{dTh1D1}
\hat{E}^{++}
\wedge
\left( \hat{E}^{2-}_{~\dot{q}} -
{\sqrt{1-F^{(0)} \over 1+ F^{(0)}}} \hat{E}^{1-}_{~\dot{q}}\right)
= 0,
\qquad
\hat{E}^{--}
\wedge
\left( \hat{E}^{2+}_{~{q}} +
{\sqrt{1+F^{(0)} \over 1 - F^{(0)}}} \hat{E}^{1+}_{~{q}}\right)
= 0,
\end{equation}
can be obtained  as a result of variations
$i_\d \hat{E}^{2-}_{~\dot{q}}$
and
$
i_\d \hat{E}^{2+}_{~{q}}$.
The independent part of the  superfield generalization of
the bosonic coordinate
equations of motion
appears as a result of the variation with respect to
$i_\d \hat{E}^I =i_\d\hat{\Pi}^{\underline{m}}
U^I_{\underline{m}}$ and is 
\begin{equation}\plabel{MID1}
M^{I}_{2(D1)} \equiv
\hat{E}^{--} \wedge f^{++I} - \hat{E}^{++} \wedge f^{--I}
-2i \g^I_{q\dot{q}}
F^{(0)} \left(\hat{E}^{2+}_{~q} \wedge \hat{E}^{2-}_{~\dot{q}} +
\hat{E}^{1+}_{~q} \wedge \hat{E}^{1-}_{~\dot{q}} \right) -
\end{equation}
$$
-2i \g^I_{q\dot{q}}
\sqrt{1-(F^{(0)})^2}
\left(\hat{E}^{2+}_{~q} \wedge \hat{E}^{1-}_{~\dot{q}} +
\hat{E}^{1+}_{~q} \wedge \hat{E}^{2-}_{~\dot{q}}\right)=0.
$$

Eq. \p{MID1} can be obtained
as  a consequence of the superembedding
equation \p{EID1c} and the superfield fermionic equations \p{dTh1D1}
with constant $F^{(0)}$.

Now it is evident that the superfield fermionic equations of motion for
the super-D1-brane \p{dTh1D1}
coincide with
the general solution
\p{dTh1a} (or \p{E-q1gs}) of the
superembedding  equation \p{EI} for
\begin{equation}\plabel{a=a(F)}
a= {\sqrt{1+F^{(0)} \over 1 - F^{(0)}}} , \qquad da=0.
\end{equation}
Note that $a$ is a constant due to the 2-dimensional
Born-Infeld equation \p{dQ0=0}.

Actually, the scale of the parameter
$a$ is unessential and  the value of the field strength of
the gauge field is in one--to--one  correspondence with
the value of an angle $\a$ which can be regarded as a parameter of the
$SO(2)$ symmetry \p{SO(2)} of the superembedding equation \p{se}
\begin{equation}\plabel{al(F)1}
Cos \a =  \sqrt{{1+ F^{(0)}\over 2 }}.
\qquad
\end{equation}
which is the flat superspace image of the classical
S-duality group $SL(2,R)$.  Thus we find that this $SO(2)$ symmetry mix
the fundamental superstring with super-D1-brane with all possible values
of constant field strength of the gauge field.

More precisely, the map \p{al(F)1} corresponds to $SO(2)$ transformations
with the parameter $\a \in [-{\pi \over 2}, {\pi \over 2}]$, while the
range $\a \in [{\pi \over 2}, {3\pi \over 2}]$, which corresponds to
the solution of superembedding equation with negative $a$,
relates superstring with anti--super--D1--brane whose action
has opposite sign of the Wess--Zumino term with respect to \p{SD1}
$S_{\bar{D}1}=S_{D1}\vert_{C_2 \rightarrow -C_2,~ C_0 \rightarrow
-C_0}$.  Thus we arrive at Eqs. \p{a(F)}, \p{al(F)}.  From the point of
view of superembedding approach to super-D1-brane, the fundamental
superstring and anti--superstring (with opposite NSNS charge
$S_{\bar{I}\bar{I}\bar{B}}= S_{IIB} \vert_{B_2 \rightarrow -B_2}$)
correspond to the non-proper limits of the value of the on--shell gauge
field strength $F^{(0)}=-1$ and $F^{(0)}=+1$.

Hence we conclude that the superembedding equation \p{EID1c},
(or, equivalently, \p{basic}) is S-duality invariant and provides a
universal description of the fundamental type IIB superstring  and
super-D1-brane.

\bigskip

\setcounter{equation}{0}

\section{Universal description of superstring and super-D1-brane
in general supergravity background.}

\subsection{Constraints and Weyl transformations}

   In general type IIB supergravity background the supervielbein
\begin{equation}\plabel{EA}
  E^{\underline{A}} =\left(  E^{\underline{a}},
  E^{\underline{\a}1}, E^{\underline{\a}2} \right)=
dZ^{\underline{M}}
  E^{~\underline{A}}_{\underline{M}}
  (Z^{\underline{M}}),
\end{equation}
    is subject to the basic torsion constraints
\cite{HW}
    \begin{equation}\plabel{IIBc}
T^{\underline{a}} \equiv {\cal D}E^{\underline{a}}
\equiv dE^{\underline{a}}- E^
{\underline{b}}\wedge w_{\underline{b}}^{~\underline{a}} =
-i
(E^{\underline{\a}1} \wedge E^{\underline{\b}1} +
E^{\underline{\a}2} \wedge E^{\underline{\b}2})
\s^{\underline{a}}_{\underline{\a}\underline{\b}}
\end{equation}
where
$${\cal D}= E^{\underline{a}}{\cal D}_{\underline{a}}+
E^{\underline{\a}1}
{\cal D}_{\underline{\a}1}
+ E^{\underline{\a}2}{\cal D}_{\underline{\a}2}
$$
is covariant external derivative and
\begin{equation}\plabel{wc}
w_{\underline{b}}^{~\underline{a}}= E^{\underline{c}}
w_{\underline{c}\underline{b}}^{~~~\underline{a}}+
E^{\underline{\a}1}w_{\underline{\a}1\underline{b}}^{~~~~\underline{a}}
+ E^{\underline{\a}2}w_{\underline{\a}2\underline{b}}^{~~~~\underline{a}},
\qquad
w^{\underline{a}\underline{b}}
=-w^{\underline{a}\underline{b}}
\end{equation}
is the $SO(1,9)$ spin connection.

The basic constraint \p{IIBc} is invariant under
the local Weyl transformations
\begin{equation}\plabel{WEac}
E^{\underline{a}} \rightarrow E^{\underline{a}\prime}  =
e^{2W} E^{\underline{a}},
\end{equation}
\begin{equation}\plabel{WEal1c}
E^{\underline{\a}1} \rightarrow E^{\underline{\a}1\prime}  =
e^{W}
\left( E^{\underline{\a}1} - i
E^{\underline{a}} \tilde{\s}_{\underline{a}}^{ \underline{\a}\underline{\b}}
\nabla_{\underline{\b}1} W
\right),
\end{equation}
\begin{equation}\plabel{WEal2c}
E^{\underline{\a}2} \rightarrow E^{\underline{\a}1\prime}  =
e^{W}
\left( E^{\underline{\a}2} - i
E^{\underline{a}} \tilde{\s}_{\underline{a}}^{ \underline{\a}\underline{\b}}
\nabla_{\underline{\b}2} W
\right),
\end{equation}
\begin{equation}\plabel{Wwab}
w^{\underline{a}\underline{b}}
\rightarrow
(w^{\underline{a}\underline{b}})^{\prime}  =
w^{\underline{a}\underline{b}} + {1\over 2} E^{[\underline{a}}
\nabla^{\underline{b}]} W + 4
   (\s^{\underline{a}\underline{b}})_{\underline{\a}}^{~\underline{\b}}
   \left( E^{\underline{\a}1} \nabla_{\underline{\b}1}W +
   E^{\underline{\a}2} \nabla_{\underline{\b}2}W \right) +
\end{equation}
$$
+ i E_{\underline{c}}
  (\tilde{\s}^{\underline{a}\underline{b}\underline{c}})
^{\underline{\a}\underline{\b}}
\left(
\nabla_{\underline{\a}1}W \nabla_{\underline{\b}1}W
+
\nabla_{\underline{\a}2}W \nabla_{\underline{\b}2}W \right)
$$
with arbitrary superfield parameter $W=W(Z)$.
It is evidently invariant as well under the local $SO(2)$ rotations of
fermionic supervielbein forms.

When superbranes in supergravity background are considered,
it is convenient to introduce
a superfield generalizations for all tensor gauge
fields involved into the  supergravity supermultiplet.
Using the string terminology, one can introduce \cite{HW,c0}
the NSNS 2-form
$B_2 =
{1 \over 2}
dZ^{\underline{M}}
\wedge
dZ^{\underline{N}}
B_{\underline{N}{\underline{M}}}(Z)$,
RR 2-form and 4-form
$C_2 =
{1 \over 2}
dZ^{\underline{M}}
\wedge
dZ^{\underline{N}}
C_{\underline{N}{\underline{M}}}(Z)$,
$C_4 =
{1 \over 4!}
dZ^{\underline{M}_4}
\wedge \ldots \wedge
dZ^{\underline{M}_1}
C_{\underline{M}_1\ldots \underline{M}_4}(Z)$, as well as an 
axion superfield $C_0(Z)$.
 The requirement of the $\kappa$--symmetry of the
 super--Dp--brane actions specifies the constraints on the NS-NS
 and RR field strengths \cite{c0}

\begin{equation}\plabel{H3c}
  H_3 \equiv dB_2 = -
i e^{{1 \over 2}\Phi} E^{\underline{a}} \wedge
(E^{\underline{\a}1} \wedge E^{\underline{\a}1} -
E^{\underline{\a}2} \wedge E^{\underline{\a}2})
\s_{\underline{a}~\underline{\a}\underline{\b}} +
\end{equation}
$$
+ {1 \over 4} e^{{1 \over 2}\Phi}  E^{\underline{b}} \wedge E^{\underline{a}}  \wedge
   \left( E^{\underline{\a}1} \nabla_{\underline{\b}1}\Phi -
   E^{\underline{\a}2} \nabla_{\underline{\b}2}\Phi\right)
   (\s_{\underline{a}\underline{b}})_{\underline{\a}}^{~\underline{\b}}
   + {1 \over 3!}
    E^{\underline{c}}  \wedge
    E^{\underline{b}}  \wedge
    E^{\underline{a}} H_{\underline{a} \underline{b}\underline{c}}
    (Z),
  $$

\begin{equation}\plabel{R1c}
 R_1 = dC_0 =
e^{- \Phi}
 E^{\underline{\a}1}
\nabla_{\underline{\a}2}\Phi
-
e^{- \Phi}
E^{\underline{\a}2}
\nabla_{\underline{\a}1}\Phi
+
E^{\underline{a}} R_{\underline{a}},
\end{equation}

 \begin{equation}\plabel{R3c}
 R_3 = dC_2 - C_0 dB_2=
2i e^{-{1 \over 2}\Phi} E^{\underline{a}}
 \wedge
{E}^{\underline{\a}2}
\wedge
{E}^{\underline{\b}1}
{\sigma}_{\underline{a}\underline{\a}\underline{\b}}
+
\end{equation}
$$
+{1 \over 4} e^{-{1 \over 2}\Phi}  E^{\underline{b}} \wedge E^{\underline{a}}  \wedge
   \left( E^{\underline{\a}1}
\nabla_{\underline{\a}2}\Phi
    +
   E^{\underline{\a}2}
\nabla_{\underline{\a}1}\Phi
  \right)
   (\s_{\underline{a}\underline{b}})_{\underline{\a}}^{~\underline{\b}}
   + {1 \over 3!}
    E^{\underline{c}}  \wedge
    E^{\underline{b}}  \wedge
    E^{\underline{a}} R_{\underline{a} \underline{b}\underline{c}}
    (Z),
  $$

 \begin{equation}\plabel{R5c}
 R_5 = dC_4 - C_2 \wedge H_3 =
2i {1 \over 3!}
E^{\underline{c}}
 \wedge
E^{\underline{b}}
 \wedge
E^{\underline{a}}
 \wedge
{E}^{\underline{\a}2}
\wedge
{E}^{\underline{\b}1}
{\sigma}_{\underline{a}\underline{b}\underline{c}~\underline{\a}\underline{\b}}
+  \ldots
\end{equation}

Actually, the constraints on the superforms of type IIB supergravity
can be written in a manifestly $SL(2,R)$ invariant manner  \cite{DLT}.
The lack of manifest $SL(2,R)$ invariance for the constraints
\p{H3c} -- \p{R5c} obtained from the consideration of
super--Dp--brane actions reflects the fact that not all the
super--Dp--branes are invariant under the S--duality.

\bigskip

\subsection{Superfield equations in general supergravity background}

In general supergravity background  the generalized actions for
superstring and super--D1--brane both
produce the superembedding equation
\begin{equation}\plabel{EIc}
\hat{E}^{I} \equiv
  \hat{E}^{\underline{a}} U^{I}_{\underline{a}}=0. \qquad
\end{equation}
The superfield fermionic equations
for type IIB superstring have the form
\begin{equation}\plabel{dTh1c} \hat{E}^{++}
 \wedge \left(\hat{E}^{-\dot{q}1} -{ i\over 8} \hat{E}^{--}
 (V\nabla_1)^{+}_{\dot{q}}\Phi \right) = 0, \qquad \end{equation}
 \begin{equation}\plabel{dTh2c}
\hat{E}^{--} \wedge  \left( \hat{E}^{2+}_{~q}
-{ i\over 8} \hat{E}^{++} (V\nabla_2)^{-}_{{q}} \Phi \right) = 0, \qquad
\end{equation}
while for the super-D1--brane  they are

 \begin{equation}\plabel{dTh1D1c}
\hat{E}^{++}
\wedge
\left( \hat{E}^{2-}_{~\dot{q}} +
{\sqrt{1-F^{(0)} \over 1+ F^{(0)}}} \hat{E}^{1-}_{~\dot{q}}\right)
- \qquad {} \qquad {}
 \end{equation}
$$
- {i \over 4}  \hat{E}^{++} \wedge \hat{E}^{--}
~V_{~\dot{q}}^{+\underline{\a}}
\left((F-{1\over 2} ) \nabla_{\underline{\a}2}\Phi
- {\sqrt{1-F^{(0)} \over 1+ F^{(0)}}}  (F+{1\over 2} )
 \nabla_{\underline{\a}1}\Phi \right) =0,
$$
\begin{equation}\plabel{dTh2D1c}
\hat{E}^{--}
\wedge \left( \hat{E}^{2+}_{~{q}} -
{\sqrt{1+F^{(0)} \over 1 - F^{(0)}}} \hat{E}^{1+}_{~{q}}\right)
+ \qquad {} \qquad {}
\end{equation}
$$
+ {i \over 4}  \hat{E}^{--} \wedge \hat{E}^{++}
~V_{~{q}}^{-\underline{\a}}
\left((F+{1\over 2} ) \nabla_{\underline{\a}2}\Phi
+ {\sqrt{1+F^{(0)} \over 1- F^{(0)}}}  (F-{1\over 2} )
 \nabla_{\underline{\a}1}\Phi \right) =0.
$$
The complete form of the bosonic coordinate equations
for the type IIB superstring ($M^I_{2(IIB)}=0$) and super--D1-brane
($M^I_{2(D1)}=0$) in general
supergravity  background
can be found in Appendix B.

The Born--Infeld equation for super--D1--brane
\p{BI}, \p{Q0} involves inputs from axion
and dilaton.
The general solution expresses the field strength $F^{(0)}$ through a
constant $c$ and images of axion and dilaton superfields on the
worldsheet superspace

\begin{equation}\plabel{F0c}
F^{(0)} (\zeta ) =
{ (c+ \hat{C}_0) e^{\hat{\Phi}}
\over \sqrt{1 +(c+ \hat{C}_0)^2 e^{2\hat{\Phi}} }}, \qquad
\hat{C}_0 \equiv C_0\left(\hat{Z}(\zeta)\right), \qquad
\hat{\Phi} \equiv \Phi \left(\hat{Z}(\zeta)\right), \qquad
c=const .
\end{equation}

The superembedding equation
\p{EIc} and the fermionic
equations of motion \p{dTh1c}, \p{dTh2c}, or
\p{dTh1D1c},
\p{dTh2D1c}
together with the Born--Infeld equation, specify the 
bosonic coordinate equations completely as they
do in the case of the flat superspace.
Thus to study the relation between super--D1--brane and superstring it is
enough to investigate the relation between the fermionic equations
\p{dTh1c}, \p{dTh2c} and
\p{dTh1D1c},
\p{dTh2D1c}.

 \subsection{Super--Weyl transformations of fermionic equations}

Let us begin with an instructive  observation about superstring
fermionic equations
\p{dTh1c}, \p{dTh2c}. They  can be written in an equivalent form
\begin{equation}\plabel{dTh1c1}
\hat{E}^{++} \wedge
\left( \hat{E}^{\underline{\a}1}
-{ i\over 8} \hat{E}^{\underline{a}}
\tilde{\s}_{\underline{a}}^{\underline{\a}\underline{\b}}
\nabla_{\underline{\b}1} \Phi
\right)
V_{\underline{\a}\dot{q}}^{~-} = 0,
 \qquad
\end{equation}
\begin{equation}\plabel{dTh2c1}
\hat{E}^{--} \wedge
\left(
\hat{E}^{\underline{\a}2}
-{ i\over 8} \hat{E}^{\underline{a}}
\tilde{\s}_{\underline{a}}^{\underline{\a}\underline{\b}}
\nabla_{\underline{\b}2} \Phi\right)
V_{\underline{\a}{q}}^{~+}  = 0, \qquad
\end{equation}
where it becomes evident that these equations can be reduced to
\p{dTh1}, \p{dTh2} by super--Weyl transformations \p{WEac}--\p{Wwab}
with the superfield parameter $W= {1 \over 8} \Phi(Z)$.

To find a proper super--Weyl transform for a simplification
of the super--D1--brane equations, it is convenient to introduce a
10--dimensional composed
scalar superfield
\begin{equation}\plabel{Fc}
F (Z ) =
{ (c+ {C}(Z)) e^{{\Phi}(Z)}
\over \sqrt{1 +(c+ {C}(Z))^2 e^{2{\Phi}(Z)} }}, \qquad
\end{equation}
which can be used to write the solution
\p{F0c} of the Born--Infeld equation \p{BI}, \p{Q0} in
the form
\begin{equation}\plabel{F0c1}
F^{(0)} (\zeta ) =
F\left(\hat{Z}(\zeta)\right), \qquad
\end{equation}

In accordance with the
constraints of type IIB supergravity
\p{R1c}, the spinor covariant derivatives of the axion
$C_0$ superfield are
expressed through the ones of the dilaton superfield $\Phi$
\begin{equation}\plabel{dC0=dP}
\nabla_{\underline{\a}1}C_0 =
e^{-\Phi}
\nabla_{\underline{\a}2}\Phi,
\qquad
\nabla_{\underline{\a}2}C_0 = -
e^{-\Phi}
\nabla_{\underline{\a}1}\Phi . \qquad
\end{equation}
Hence, the spinor covariant derivatives of the superfield $F$
\p{Fc} have the form of $SO(2)$ transformed
Grassmann derivatives of dilaton superfield times
a local scale
\begin{equation}\plabel{dF=dP}
{1 \over 1-F^2} \nabla_{\underline{\a}1} F =
F \nabla_{\underline{\a}1}\Phi  +
\sqrt{1-F^2} \nabla_{\underline{\a}2}\Phi ,
\end{equation}
$$
{1 \over 1-F^2} \nabla_{\underline{\a}2} F =
- \sqrt{1-F^2}\nabla_{\underline{\a}1}\Phi  +
F \nabla_{\underline{\a}2}\Phi .
$$

Taking into account Eqs. \p{dF=dP} and performing some straightforward
calculations one can find that the super--Weyl transformation
\p{WEac}--\p{Wwab} with the parameter
\begin{equation}\plabel{WD1}
e^W =
\left( e^{-2 \Phi} + (c+C_0)^2 \right)^{{1 \over 8}}
\end{equation}
(supplemented with a trivial rescaling of the equations)
can be used to recast  Eqs.  \p{dTh1D1c}, \p{dTh2D1c}
to

 \begin{equation}\plabel{dTD1c1}
\hat{E}^{++}
\wedge
\Big(
\sqrt{{1-F^{(0)}\over 2}}
\left(
\hat{E}^{\underline{\a}1}
-{ i\over 8} \hat{E}^{\underline{a}}
\tilde{\s}_{\underline{a}}^{\underline{\a}\underline{\b}}
\nabla_{\underline{\b}1} \Phi
\right) +
 \qquad {} \qquad
 \end{equation}
$$
{} \qquad {} \qquad + \sqrt{{1+F^{(0)}\over 2}}
\left(
\hat{E}^{\underline{\a}2}
-{ i\over 8} \hat{E}^{\underline{a}}
\tilde{\s}_{\underline{a}}^{\underline{\a}\underline{\b}}
\nabla_{\underline{\b}2} \Phi
\right)
\Big)
V_{\underline{\a}\dot{q}}^{~-} = 0,
$$

\begin{equation}\plabel{dTD1c2}
\hat{E}^{--}
\wedge
\Big(
- \sqrt{{1+F^{(0)}\over 2}}
\left(
\hat{E}^{\underline{\a}1}
-{ i\over 8} \hat{E}^{\underline{a}}
\tilde{\s}_{\underline{a}}^{\underline{\a}\underline{\b}}
\nabla_{\underline{\b}1} \Phi
\right)+
 \qquad {} \qquad 
\end{equation}
$$
{} \qquad {} \qquad +
\sqrt{{1-F^{(0)}\over 2}}
\left(
\hat{E}^{\underline{\a}2}
-{ i\over 8} \hat{E}^{\underline{a}}
\tilde{\s}_{\underline{a}}^{\underline{\a}\underline{\b}}
\nabla_{\underline{\b}2} \Phi
\right)
\Big)
V_{\underline{\a}q}^{~+} = 0.
$$

Eqs. \p{dTD1c1}, \p{dTD1c2} are related with the superstring equations
\p{dTh1c}, \p{dTh2c} by $SO(2)$ rotations with the parameter
\begin{equation}\plabel{al(F)c}
Cos ~\a = \sqrt{{1+ F(Z)\over 2 }}.
\qquad
\end{equation}

So, in general supergravity background the superstring and
super--D1--brane are related by $SO(2)$ transformations whose
parameter is constructed from the
$10$--dimensional counterpart $F(Z)$ of the
generalized field strength $F^{(0)}$ of the super--D1--brane gauge field
in the same way as in the flat superspace.  However, in
distinction to the case of flat superspace, $F^{(0)}$
is not constant, but is constructed from images of the axion and dilaton
superfields on the worldsheet superspace \p{F0c}.
Thus the superstring
and super--D1--branes are related by a {\sl local } $SO(2)$ rotations, but
with the parameter \p{al(F)c} dependent on the  superspace
coordinates through the mediation of the axion and dilaton superfield
only \p{Fc}.  Such rotation appears as a compensated $SO(2)$ rotation
\cite{CWZ,DV}
when the $SL(2,R)$ (classical S--duality group) acts on the matrix
constructed form the axion and dilaton
\begin{equation}\plabel{K} K =
e^{{1\over 2} \Phi } \left(\matrix{ e^{-\Phi} & C_0 \cr 0     & 1  \cr }
\right) \quad \in \quad {SL(2,R) \over SO(2)}
\qquad
\end{equation}
Indeed, multiplying the
matrix \p{K} by an element of global $SL(2,R)$ group
\begin{equation}\plabel{G}
G =
\left(\matrix{
{1 \over c} \Big({1 \over \gamma } + \b \Big) & \b \cr
                                           \g &  c\g \cr }
		   \right)
\quad \in \quad SL(2,R)
\end{equation}
from the left, one arrives at the transformation low for axion and dilaton
superfield
\begin{equation}\plabel{GK=KH}
G K(\Phi, C_0) =
K(\Phi^\prime , C_0^\prime )
H(G, \Phi, C_0),
\qquad
\end{equation}
where
\begin{equation}\plabel{H}
H(G, \Phi, C_0)
 =
\left(\matrix{ Cos~2\a & Sin~ 2 \a  \cr
	       -Sin 2\a & Cos~2\a \cr}
		   \right)
\quad  \in \quad SO(2), 
\end{equation}
\begin{equation}\plabel{Cos2a}
Cos~2\a = -
{ (c+ {C}(Z)) e^{{\Phi}(Z)}
\over \sqrt{1 +(c+ {C}(Z))^2 e^{2{\Phi}(Z)} }}
= - F (Z ). \qquad
\end{equation}
Eqs. \p{GK=KH}, \p{H}, \p{Cos2a} describe an induced $SO(2)=U(1)$
transformations acting on fields {\sl with $U(1)$ charge equal to 2}.  The
$SO(2)$ rotation matrix acting on a doublet of fields which can be
regarded as a complex field with $U(1)$ charge 1 has the form
\begin{equation}\plabel{SO(2)c1} \left(\matrix{ Cos~ \a & Sin~ \a \cr -
Sin~ \a & Cos~ \a \cr } \right) \quad \in \quad SO(2) \end{equation} with
\begin{equation}\plabel{al(F)c1} Cos ~\a = \sqrt{{1+ Cos ~2\a \over 2 }}=
\sqrt{{1 - F(Z)\over 2 }}.
\qquad
\end{equation}
The matrix \p{SO(2)c1} describes  $SO(2)$ rotations
of the fermionic
supervielbein forms (cf. \cite{DLT})
 \begin{equation}\plabel{SO(2)c}
\left(\matrix{ E^{\underline{\a}1 \prime} \cr
E^{\underline{\a}2\prime} } \right)=
\left(\matrix{ Cos~ \a & Sin~ \a \cr
- Sin~ \a & Cos~ \a \cr }
\right)
\left(\matrix{ E^{\underline{\a}1} \cr
\hat{E}^{\underline{\a}2} } \right)
\end{equation}
and can be used to relate the super--D1--brane and superstring in the same
manner as in the flat superspace.

\bigskip

To pass
from  super--D1--brane equations \p{dTD1c1}, \p{dTD1c2}
with the value of the field strength \p{F0c}
to the superstring ones \p{dTh1c1}, \p{dTh2c1}
we need in the super--Weyl transformations \p{WD1} as well.
They actually contain an important information.
Indeed, if one performs such super--Weyl transformations in an
action of 2--dimensional supersymmetric object in supergravity
background, e.g.  in the functional \p{SD1}
taken on the surface of Born--Infeld equations \p{Fc}, one arrives at the
effective tension renormalized as follows
\begin{equation}\plabel{T1}
T^\prime (Z)= T \sqrt{
e^{-2 \Phi} + (c+C_0)^2}.
\end{equation}
If one considers Eq. \p{T1} in a constant
axion and dilaton background
$$\Phi = <\Phi > = const, \qquad C_0 = <C_0 > =
const $$
and remembers that the Dirac quantization condition requires
that both constants $T$ and $cT$ are integer valued
(see e.g. \cite{LLPS})
\begin{equation}\plabel{TpTq}
 T= p ~~\in ~~
{\hbox{\bf Z}}, \qquad cT= q ~~\in ~~ {\hbox{\bf Z}}, \qquad
\end{equation}
 one arrives at the famous formula for the $(p,q)$ string tension
 \cite{Schwarz}
 \begin{equation}\plabel{Tpq}
 T_{p,q} = \sqrt{ p^2e^{-2
\Phi} + (q+pC_0)^2}.  \end{equation}

Thus we can state that in general type IIB supergravity background

\begin{itemize}
\item
superstring and super--D1--brane are described in a universal manner by
the superembedding equation \p{EIc}.

\item
They are related by the super--Weyl transformations \p{WD1}  and induced
$SO(2)$ rotations \p{SO(2)c}, whose parameter \p{al(F)c1} is dependent on
the worldsheet superspace coordinates through the mediation of the
axion and dilaton superfields only \p{Fc}. 
They are  the compensating
$SO(2)$ rotations \cite{CWZ,DV} for  $SL(2,R)$ transformation \p{G} acting
on the axion and dilaton superfields \p{GK=KH}.

\item
Superstring equations appear as a singular limit $F^{(0)} \rightarrow -1$
of the super--D1--brane equations.

\item
The super--Weyl transform \p{WD1} reproduces the correct formula for
the $(p,q)$ string tension \p{Tpq}.

\end{itemize}

\bigskip

\section{Conclusion and outlook}

In this paper we worked out the generalized action principle
for the fundamental superstring and
super-D1-brane and derived the superfield equations of motion.
In such a way we  have proved that the superembedding equation \p{EI}
(or, equivalently, \p{basic}) which describes an embedding of
the worldsheet superspace $\S^{(2|8+8)}$ with 2 bosonic and 16 fermionic
directions into the $D=10$ type IIB target superspace
${\underline{{\cal M}}}^{(10|16+16)}$,
provides a universal, S-duality invariant description of
the fundamental type IIB superstring
and Dirichlet superstring (super-D1-brane).

In the case of flat target superspace
the S-duality transformations
manifest themselves
in the $SO(2)$ symmetry of the superembedding equation.
This continuous symmetry and the Weyl rescaling
mix the fundamental superstring
with a set of super-D1-branes 'marked' by a constant value of the
(on-shell)
field strength $-1 \leq F^{(0)}\leq 1$ of the worldvolume gauge field.
The superstring with unit NSNS charge
corresponds to the limiting value $F^{(0)}=- 1$ of the gauge field
strength.

We studied as well the case of general supergravity background
and find that the $SO(2)$ rotation related the super--D1--brane and
superstring has a parameter determined by the value of a constant and the  
dilaton and axion superfields. This is a compensating $SO(2)$ rotation for
the S-duality $SL(2,R)$ transformation acting on $2 \times 2$ unimodular
matrix constructed from the axion and dilaton superfields.
The super--Weyl transform, which should be used together with the $SO(2)$
rotation on the way from the super--D1--brane to superstring, allows to
reproduce the formula for $(p,q)$ string tension \p{Tpq} \cite{Schwarz}.
This supports the conclusion
\cite{HK} that the   $(p,q)$ string
can be associated with
a D-brane action
considered on the shell of Born--Infeld equation
\cite{HK}.

Note that one can not see any non--Abelian structure in such description,
as it includes the description of just
the bound state
of $p$ fundamental strings and $q$ super-D1-branes,
but not of the excitations over such  bound state, which
are described by non-Abelian $SU(q)$ SYM theory in the linearized
approximation
\cite{Witten}.
We hope that en effective  Lagrangian
description of a system of $q$ coincident super-Dp-branes
interacting through an exchange by fundamental superstrings
can be found by working out the  recently proposed approach to the action
of interacting superbrane systems \cite{BK}.

The similar situation must occur for the superembedding equation,
describing an embedding of the worldsheet superspace
$\S^{(6|16)}$ with 6 bosonic and 16 fermionic
directions into $D=10$ type IIB target superspace
${\underline{{\cal M}}}^{(10|16+16)}$.
It should provide a universal description of
the super--D5--brane, type IIB super--NS5--brane and super--KK5--brane
(a D=10 type IIB Kaluza--Klein monopole).
Such a study is of a special interest because the complete
supersymmetric action are still unknown for the latter
objects (see \cite{Lozano} and refs. therein for the bosonic actions).

We conclude with the following observation.
Let us
consider a system of several
non-intersecting super-D1-branes and
fundamental superstrings. Such  system certainly preserves supersymmetry and,
thus can be described in term of worldsheet superspaces.
It is intriguing that we can consider them as one, but multiply connected
worldsheet superspace characterized by one universal superembedding equation
\p{basic} or, equivalently, \p{EI}.
Such  superspace has several connected components
$$
\S^{(2|16)}
 = \oplus_{k}
 \S^{(2|16)}_k.
$$
The solution of the Born-Infeld equation \p{dQ0=0}
$dF^{(0)}=0$ (or, more generally, \p{BI}) provides each of the connected
components $\S^{(2|16)}_k$ with a specific value of the constant field
strength of the worldvolume gauge field $F^{(0)}_k$
(or $c_k$ \p{Fc}).

Thus one superembedding equation describes actually the system of several
non-intersecting super-D1-branes  and superstrings on the mass shall.
The number of branes coincides with the number of
connected components the worldvolume superspace is split on.

It is quite interesting to try to apply the superembedding approach
for the system of several interacting branes: intersecting branes and/or
open branes ending on other branes (see \cite{HSopen} for some results
in this directions). However the problem one immediately meets
is as follows.
The language of worldvolume superspace used by the superembedding
approach is certainly proper for a system which preserves supersymmetry.
And this is not the case for the general system of interacting branes.

One of the possible ways is to develop the superembedding approach
for the systems with  'soft supersymmetry breaking', e.g. to incorporate
somehow
terms dependent explicitly on the Grassmann coordinate of the
worldvolume superspace into the superembedding equations.
On the other hand,
the most physically interesting systems, which consist of several
coincident D-branes interacting through exchange by fundamental
superstrings, preserve the same amount of supersymmetry as the free
branes.  So, hopefully, one can find a description of such a system in
a framework of some non-Abelian modification of the superembedding
approach\footnote{An existence of such approach is indicated indirectly
by recent results on an existence
of a non--Abelian generalization of the $\kappa$--symmetry
\cite{Bergshoeff}. Note also a search for a noncommutative generalization of 
the background geometry \cite{S00}.}.

\section*{Acknowledgments}

The author is grateful to
D. Sorokin and M. Tonin for interest in
this work and  useful conversations. He thanks the Padova Section of INFN
for the hospitality at the Padova University
where a part of this work
has been done.
A partial support from Ukrainian GKNT grant {\bf 2.5.1/52} is acknowledged.

\renewcommand\baselinestretch{1.0}

\setcounter{equation}{0}
\def\theequation{A.\arabic{equation}}

\newpage
\section*{Appendix A:  Lorentz harmonics and equivalent forms of
superembedding equations}

As it was pointed out in Section 1.1, the superembedding equations
\p{basic} can be represented in the differential form notations
as Eq. \p{se} \cite{bpstv}
\begin{equation}\plabel{se1}
\hat{E}^{\underline{a}}\equiv
 d{Z}^{\underline{M}}(\zeta )
\hat{E}^{~\underline{a}}_{\underline{M}}
 \left(\hat{Z}^{\underline{M}}(\zeta)\right) =
e^{++} \hat{E}^{~\underline{a}}_{++} +
e^{--} \hat{E}^{~\underline{a}}_{--}.
\end{equation}
The bosonic forms $e^{\pm\pm}$  of the worldsheet
supervielbein \p{aA} have not been specified by any conditions. Hence,
 the components $\hat{E}^{\underline{a}}_{\pm\pm}$ have not been
 specified as well.  The arbitrariness in the choice of $e^{\pm\pm}$
 includes, in particular, general linear transformations $$ e^{++\prime} =
\a e^{++} + \b_{--}^{++} e^{--}, \qquad e^{--\prime} = \b_{++}^{--} e^{++}
+ \rho e^{--}, \qquad det\left( \matrix{ \a   & \b_{--}^{++}  \cr
\b_{++}^{--} & \rho \cr }\right) \not= 0, $$ which imply the local
$GL(2,R)$ redefinition of the 10-vector fields
$\hat{E}^{\underline{a}}_{\pm\pm}(\zeta )$

 \begin{equation}\plabel{GL}
\a \hat{E}^{\underline{a}\prime}_{++} +
\b_{++}^{--}\hat{E}^{\underline{a}\prime}_{--}
= \hat{E}^{\underline{a}}_{++}, \qquad
\b_{--}^{++}\hat{E}^{\underline{a}\prime}_{++} +
\rho \hat{E}^{\underline{a}\prime}_{--}
= \hat{E}^{\underline{a}}_{--}. \qquad
\end{equation}
These transformations can be used
to achieve the light--likeness of both
$\hat{E}^{\underline{a}}_{++}(\zeta )$ and
$\hat{E}^{\underline{a}}_{--}(\zeta )$
\begin{equation}\plabel{Virasoro}
\hat{E}^{\underline{a}}_{++}
\eta_{\underline{a}\underline{b}}\hat{E}^{\underline{b}}_{++}
=0, \qquad
\hat{E}^{\underline{a}}_{--}
\eta_{\underline{a}\underline{b}}\hat{E}^{\underline{b}}_{--}
=0, \qquad
\end{equation}
as well as normalization conditions
\begin{equation}\plabel{E+E-}
\hat{E}^{\underline{a}}_{++}
\eta_{\underline{a}\underline{b}}
\hat{E}^{\underline{b}}_{--}
={1 \over 2}.  \qquad
\end{equation}

The set of two light-like normalized vectors
always can be completed up to the
$SO(1,9)$ valued matrix
\begin{equation}\plabel{vhII0}
U^{(\underline{b})}_{\underline{a}}
=
\left( U^{0}_{\underline{a}},
U^{~J}_{\underline{a}},
U^{9}_{\underline{a}}\right)
=
\left(
(\hat{E}_{--\underline{a}}+\hat{E}_{++\underline{a}}),
U^{~J}_{\underline{a}},
(\hat{E}_{--\underline{a}}-\hat{E}_{++\underline{a}})
\right)
~ \in ~ SO(1,9).
\end{equation}
To this end one introduces a set of $8$ independent vectors
$U^{\underline{a}I}$, $I=1,\ldots 8$ which are orthogonal to
$\hat{E}^{\pm\pm}_{\underline{a}}$ and normalized
\begin{equation}\plabel{UI}
 \hat{E}_{\pm\pm}^{\underline{a}} U_{\underline{a}}^I=0, \qquad
U^{\underline{a}I}U_{\underline{a}}^J= - \d^{IJ}.
\end{equation}

For convenience we can denote
\begin{equation}\plabel{E=U}
\hat{E}^{\underline{a}\prime}_{++}
\equiv  { 1\over 2} U^{\underline{a}--}, \qquad
\hat{E}^{\underline{a}\prime}_{--}
\equiv { 1\over 2} U^{\underline{a}++}.  \qquad
\end{equation}
Than Eq. \p{vhII0} acquires universal form \p{vhII}
\begin{equation}\plabel{vhII1}
U^{(\underline{b})}_{\underline{a}}
=
\left( U^{0}_{\underline{a}},
U^{~J}_{\underline{a}},
U^{9}_{\underline{a}}\right)
=
\left({1 \over 2} (U^{++}_{\underline{a}}+U^{--}_{\underline{a}}),
U^{~J}_{\underline{m}},
{1 \over 2} (U^{++}_{\underline{a}}-U^{--}_{\underline{a}})
\right)
~ \in ~ SO(1,9).
\end{equation}

\bigskip

\setcounter{equation}{0}
\def\theequation{B.\arabic{equation}}

\section*{Appendix B:
On Superstring and super-D1-branes in curved superspace
}

Here we collect some useful formulae for superstring and super--D1--brane
in general $D=10$ type IIB supergravity background.

\subsection*{B1. Cartan forms and Maurer--Cartan equations in curved
superspace}

 Cartan forms  \p{fpmpmI}, \p{omAIJ}
are invariant under the
{\sl global} Lorentz transformations of the
vectors $U^{\pm\pm}_{\underline{a}}, U^{I}_{\underline{a}}$.
However, when the superstring in curved superspace is considered,
the harmonics are transformed by the {\sl local} Lorentz
group $SO(1,9)$
\begin{equation}\plabel{SO(1,9)U}
SO(1,9): \qquad {} \qquad
U^{\underline{a}(\underline{b})\prime} (\zeta )
=
U^{\underline{c}(\underline{b})} (\zeta )
L_{\underline{c}}^{~\underline{a}} (\hat{Z}(\zeta)). \qquad
\end{equation}
whose action on the supervielbein of
type  IIB supergravity is defined by
\begin{equation}\plabel{SO(1,9)Ev}
E^{\underline{a}\prime}= E^{\underline{b}}(Z)
L_{\underline{b}}^{~\underline{a}} (Z), \qquad {} \qquad
L_{\underline{c}}^{~\underline{a}} \eta^{\underline{c}\underline{d}}
L_{\underline{d}}^{~\underline{b}}= \eta^{\underline{a}\underline{b}},
\end{equation}
\begin{equation}\plabel{LsL=sL}
E^{\underline{\a}1,2 \prime}= E^{\underline{\b}1,2}(Z)
{\cal L}_{\underline{\b}}^{~\underline{\a}} (Z), \qquad  {} \qquad
{\cal L}_{\underline{\g}}^{~\underline{\a}}
\tilde{\s}^{\underline{a}}
{}_{\underline{\a}\underline{\b}}
{\cal L}_{\underline{\d}}^{~\underline{\b}}=
{\s}^{\underline{b}}{}_{\underline{\g}\underline{\d}}
L_{\underline{b}}^{~\underline{a}} (Z). \qquad
\end{equation}

Under the local $SO(1,9)$ group \p{SO(1,9)U} the
Cartan forms \p{omAIJ}, \p{fpmpmI}
are transformed  inhomogeneously. If we collect them in the
antisymmetric (Lorentz algebra valued) matrix
\begin{equation}\plabel{Cfst}
 \Omega^{\underline{a}\underline{b}}
  \equiv
U^{\underline{a}}_{\underline{m}} d U^{\underline{b}\underline{m}}
 = \left( \matrix{
 0 & { f^{++J} + f^{--J}\over 2} & -{1 \over 2}\omega   \cr
-{ f^{++I} + f^{--I}\over 2} &  A^{IJ} & - { f^{++I} - f^{--I}\over 2} \cr
  {1 \over 2}\omega  &  { f^{++J} - f^{--J}\over 2}  & 0 \cr }  \right)
\end{equation}
then the transformation low can be written in a compact form
\begin{equation}\plabel{SO(1,9)Om}
SO(1,9): \qquad {} \qquad
\Om^{(\underline{a})(\underline{b})\prime}
=
\Om^{(\underline{a})(\underline{b})} +
U^{\underline{c}(\underline{a})}
(L^{-1}dL)_{\underline{c}}^{~\underline{d}}
U_{\underline{d}}^{(\underline{b})}.
\end{equation}

To construct the forms which are invariant under \p{SO(1,9)U} one can
add to the original definition \p{Cfst} the supergravity spin connections 
$w_{\underline{a}}^{~\underline{b}}$ contracted with the harmonic vectors
\begin{equation}\plabel{tOm}
\tilde{\Om}^{(\underline{b})(\underline{a})}=
\Om^{(\underline{a})(\underline{b})}
+ U^{(\underline{a})\underline{c}} w^{~\underline{d}}_{\underline{c}}
U^{(\underline{b})}_{\underline{d}}=
U^{(\underline{a})\underline{c}}
\left(d U^{(\underline{b})}_{\underline{c}} +
w^{~\underline{d}}_{\underline{c}}
U^{(\underline{b})}_{\underline{d}}\right)
\end{equation}
$$
SO(1,9): \qquad {} \qquad
\tilde{\Om}^{(\underline{b})(\underline{a})\prime}
= \tilde{\Om}^{(\underline{b})(\underline{a})}.
$$
(Actually this is a prescription for the construction of the so called
'gauge fields of nonlinear realization' \cite{CWZ}).

Hence, the local Lorentz invariant  coset vielbeine,
$SO(1,1)$ and $SO(8)$ connections are
\begin{equation}\plabel{fI}
\tilde{f}^{\pm\pm I} \equiv f^{\pm\pm I} +
(UwU)^{\pm\pm I} \equiv
U^{\pm\pm}_{\underline{a}}
(d U^{I\underline{a}} + w^{~\underline{b}}_{\underline{a}}
U_{\underline{b}}^{I}),
\end{equation}
\begin{equation}\plabel{tom}
 \tilde{\om}
  \equiv  \om + {1 \over 2} (UwU)^{-- | ++}=
{1 \over 2}
U^{--\underline{a}} (d U^{++}_{\underline{a}}+
w^{~\underline{b}}_{\underline{a}}
U^{++}_{\underline{b}}), \qquad
\end{equation}
\begin{equation}\plabel{tAIJ}
 \tilde{A}^{IJ} = {A}^{IJ} + (UwU)^{IJ}
  \equiv
U^{I\underline{a}} (d U^{J}_{\underline{a}}
+w^{~\underline{b}}_{\underline{a}}
U^{J}_{\underline{b}}), \qquad
\end{equation}
where $f^{\pm\pm I}, A^{IJ}, \om$ are the original Cartan forms
\p{fpmpmI}, \p{omAIJ}.

The worldsheet covariant derivatives and, thus,
the worldsheet superspace torsion can be defined with the use of
the connections \p{tom}, \p{tAIJ} induced by embedding
\begin{equation}\plabel{t++d}
t^{++}=De^{++} \equiv de^{++} - e^{++} \wedge \tilde{\om},
\end{equation}
\begin{equation}\plabel{t--d}
t^{--}=De^{--} =  de^{--} + e^{--} \wedge \tilde{\om},
\end{equation}
\begin{equation}\plabel{t+qd}
t^{+q}=De^{+q} = de^{+q} -{ 1 \over 2} e^{+q} \wedge \tilde{\om} +
e^{+p} \wedge \tilde{A}^{pq},
\end{equation}
\begin{equation}\plabel{t-qd}
t^{-\dot{q}}=De^{-\dot{q}} =
de^{-\dot{q}} + { 1\over 2} e^{-\dot{q}}\wedge \tilde{\om}+
e^{-\dot{p}} \wedge \tilde{A}^{\dot{p}\dot{q}}
\end{equation}
where $\tilde{A}^{pq}$ and $\tilde{A}^{\dot{p}\dot{q}}$
are s- and c-spinor representations for $SO(8)$ connection form
$\tilde{A}^{IJ}$
\begin{equation}\plabel{Apq}
\tilde{A}^{pq} = {1 \over 4} \tilde{A}^{IJ} \g^{IJ}{}_{pq},
\qquad
\tilde{A}^{\dot{p}\dot{q}} = {1 \over 4} \tilde{A}^{IJ}
\tilde{\g}^{IJ}{}_{\dot{p}\dot{q}}.
\end{equation}
In \p{Apq} $\g^I_{p\dot{q}}\equiv \tilde{\g}^I_{\dot{q}p}$ are
$SO(8)$ Klebsh--Gordan coefficients
\begin{equation}\plabel{gI}
\g^I_{q\dot{q}}\tilde{\g}^J_{\dot{q}p} +
\g^J_{q\dot{q}}\tilde{\g}^I_{\dot{q}p} = \d^{IJ}\d_{qp}, 
\qquad {} \qquad 
\tilde{\g}^I_{\dot{q}p} \g^J_{p\dot{p}}+
\tilde{\g}^J_{\dot{q}p} \g^I_{p\dot{p}}
= \d^{IJ}\d_{\dot{q}\dot{p}},
\end{equation}
$$
\g^{IJ}_{qp}={ 1\over 2} (
\g^I\tilde{\g}^J- \g^J \tilde{\g}^I)_{qp}, \qquad {} \qquad 
\tilde{\g}^{IJ}_{\dot{q}\dot{p}} =
 { 1\over 2} (
\tilde{\g}^I\g^J- \tilde{\g}^J\g^I)_{\dot{q}\dot{p}}.
$$

The $SO(1,9)_L \otimes [SO(1,1)\times SO(8)]_R$
covariant derivatives of the
harmonic variables are defined with $SO(1,1)$ and $SO(8)$ connections
and are expressed through covariant forms $\tilde{f}^{\pm\pm I}$

\begin{equation}\plabel{DU++}
{\cal D}U^{++}_{\underline{a}}\equiv
 dU^{++}_{\underline{a}}- U^{++}_{\underline{a}} \tilde{\om }
+ w_{\underline{a}}^{~\underline{b}} U^{++}_{\underline{b}} =
U^{I}_{\underline{a}}\tilde{f}^{++I}
\end{equation}
\begin{equation}\plabel{DU--}
{\cal D}U^{--}_{\underline{a}}\equiv
 dU^{--}_{\underline{a}}+ U^{--}_{\underline{a}} \tilde{\om }
+ w_{\underline{a}}^{~\underline{b}} U^{--}_{\underline{b}} =
U^{I}_{\underline{a}}\tilde{f}^{--I}
\end{equation}
\begin{equation}\plabel{DUI}
{\cal D}U^{I}_{\underline{a}}\equiv
 dU^{I}_{\underline{a}}+ U^{J}_{\underline{a}} \tilde{A}^{JI}
+ w_{\underline{a}}^{~\underline{b}} U^{I}_{\underline{b}} =
{1 \over 2} U^{--}_{\underline{a}} \tilde{f}^{++I}
+ {1 \over 2} U^{++}_{\underline{a}} \tilde{f}^{--I}.
\end{equation}

The covariant derivatives of spinor harmonics are

\begin{equation}\plabel{DV+}
{\cal D}V^{~+}_{\underline{\a}q}\equiv
dV^{~+}_{\underline{\a}q} -
V^{~+}_{\underline{\a}q} \tilde{\om } +
V^{~+}_{\underline{\a}p} \tilde{A}^{pq}
+ w_{\underline{\a}}^{~\underline{\b}} V^{~+}_{\underline{\b}q}=
{1 \over 2} \tilde{f}^{++I}  \g^{I}_{q \dot{q}}
V^{~-}_{\underline{\a}\dot{q}}
\end{equation}

\begin{equation}\plabel{DV-}
{\cal D}V^{~-}_{\underline{\a}\dot{q}}\equiv
dV^{~-}_{\underline{\a}\dot{q}} +
V^{~-}_{\underline{\a}\dot{q}} \tilde{\om } +
V^{~-}_{\underline{\a}\dot{p}} \tilde{A}^{\dot{p}\dot{q}}
+ w_{\underline{\a}}^{~\underline{\b}} V^{~-}_{\underline{\b}\dot{q}}=
{1 \over 2} \tilde{f}^{--I}
V^{~+}_{\underline{\a}{q}} \g^{I}_{q \dot{q}}.
\end{equation}

The integrability conditions
for Eqs. \p{DU++}, \p{DU--}, \p{DUI}
indicate that the covariant Cartan forms satisfy the Maurer-Cartan equation
\begin{equation}\plabel{MC}
d\tilde{\Om}^{(\underline{a})(\underline{b})}-
\tilde{\Om}^{(\underline{a})(\underline{c})}\wedge
\tilde{\Om}_{(\underline{c})}^{~(\underline{b})}=
  U^{\underline{c}(\underline{a})} R_{\underline{c}}^{~\underline{d}}
U^{(\underline{b})}_{\underline{d}}.
\end{equation}
It can be split on the following equations for the
forms \p{fI}, \p{tom}, \p{tAIJ}
\begin{equation}\plabel{Dtf+}
{\cal D}\tilde{f}^{++I}\equiv  d\tilde{f}^{++I} - \tilde{f}^{++I}\wedge
\tilde{\om} + \tilde{f}^{++J} \wedge A^{IJ} =
  U^{\underline{a}++} R_{\underline{a}}^{~\underline{b}}
U^{I}_{\underline{b}},
\end{equation}
\begin{equation}\plabel{Dtf-}
{\cal D}\tilde{f}^{--I}\equiv  d\tilde{f}^{--I} + \tilde{f}^{--I}\wedge
\tilde{\om} + \tilde{f}^{--J} \wedge A^{IJ} =
  U^{\underline{a}--} R_{\underline{a}}^{~\underline{b}}
U^{I}_{\underline{b}},
\end{equation}
\begin{equation}\plabel{dtom}
d\tilde{\om}= {1 \over 2}
\tilde{f}^{--I}\wedge \tilde{f}^{++I} +
 {1 \over 2} U^{\underline{a}--} R_{\underline{a}}^{~\underline{b}}
U^{++}_{\underline{b}},
\end{equation}
\begin{equation}\plabel{dtA}
d\tilde{A}^{IJ} + \tilde{A}^{IK} \wedge \tilde{A}^{KJ} =
\tilde{f}^{--[I}\wedge \tilde{f}^{++J]} +
 U^{\underline{a}I} R_{\underline{a}}^{~\underline{b}}
U^{J}_{\underline{b}}.
\end{equation}

\subsection*{B2. Superstring}

The derivative of the Lagrangian form of the fundamental superstring
in general type IIB supergravity background  is
\begin{equation}\plabel{dLIIBc}
d\hat{{\cal L}}^{IIB}_2 =
2i e^{{1 \over 2} \hat{\Phi}} \hat{E}^{++} \wedge  \hat{E}^{-\dot{q}1} 
\wedge  \hat{E}^{-\dot{q}1}
- 2i e^{{1 \over 2} \hat{\Phi}} \hat{E}^{++}
\wedge  \hat{E}^{+{q}2} \wedge  \hat{E}^{+{q}2} +
\end{equation}
$$
+ {1 \over 2} e^{{1 \over 2} \hat{\Phi}}   \hat{E}^{++} \wedge
\hat{E}^{--} \wedge
\left(  \hat{E}^{1-}_{~\dot{q}} (V\nabla_1)^+_{\dot{q}} \Phi
+ \hat{E}^{2+}_{~{q}} (V\nabla_2)^-_{{q}} \Phi
\right)
+ {1 \over 2} e^{{1 \over 2} \hat{\Phi}} \hat{E}^I \wedge M^I_2 +
\propto \hat{E}^I \wedge \hat{E}^J.$$
where $\hat{M}^I_{2~IIB}$ is the l.h.s. of
the bosonic coordinate equation of motion for superstring in general type IIB
supergravity background

\begin{equation}\plabel{MIc}
\hat{M}^I_{2~IIB} \equiv
\hat{E}^{--} \wedge
\tilde{f}^{++I} - \hat{E}^{++} \wedge
\tilde{f}^{--I}
+ 4i \left(\hat{ E}^{+q1} \wedge \hat{E}^{-\dot{q}1} -
\hat{E}^{+2}_{q} \wedge \hat{E}^{-2}_{\dot{q}} \right) \gamma^I_{q\dot{q}} -
\end{equation}
$$
- \hat{E}^{++} \wedge
\left( \hat{E}^{-\dot{q}1} {\g}^I_{q\dot{q}}
(V\nabla_1)^-_{{q}} \Phi -
 \hat{E}^{-\dot{q}2} {\g}^I_{q\dot{q}}
(V\nabla_2)^-_{{q}} \Phi\right) -
$$
$$
- \hat{E}^{--} \wedge
\left( \hat{E}^{+{q}1} {\g}^I_{q\dot{q}}
(V\nabla_1)^-_{\dot{q}} \Phi -
 \hat{E}^{+{q}2} {\g}^I_{q\dot{q}}
(V\nabla_2)^+_{\dot{q}} \Phi \right) +
$$
$$
+ {1 \over 2} \hat{E}^{++} \wedge \hat{E}^{--}
\left( H_{\underline{a}\underline{b}\underline{c}}
 U^{++\underline{a}}  U^{--\underline{b}} U^{I\underline{c}}
e^{- {1\over 2}\Phi} - (U\nabla )^I \Phi
\right)=0.
$$
Here
$$
(U\nabla)^{I}
\equiv U^{I\underline{a}}
\nabla_{\underline{a}} \equiv
U^{I\underline{a}}
{E}_{\underline{a}}^{\underline{M}} (\hat{Z}) \partial_{\underline{M}},
\qquad
$$
$$
(V\nabla_2)^{+}_{\dot{q}}
\equiv V^{+\underline{\a}}_{\dot{q}}
\nabla_{\underline{\a}2} \equiv V^{-\underline{\a}}_{\dot{q}}
{E}_{\underline{\a}2}^{\underline{M}} (\hat{Z}) \partial_{\underline{M}},
\qquad
(V\nabla_1)^{-}_{{q}}
\equiv V^{-\underline{\a}}_{{q}}
\nabla_{\underline{\a}1},
$$
and $\tilde{f}^{\pm\pm I}$ are Cartan forms \p{fI}.

\bigskip

\subsection*{B3. Super--D1--brane}

After some algebraic manipulations (but without any use
of the supergravity constraints) one can present
the external derivative of the Lagrangian 2-form \p{SD1} as

\begin{equation}\plabel{dL2D1}
d\hat{{\cal L}}^{D1}_2 =
+ {e^{-\hat{\Phi}} \over
\sqrt{1-(F^{(0)})^2}} d\hat{{\cal L}}_2^{IIB} +
\hat{R}_3 +
 e^{-\hat{\Phi}} \sqrt{1+F^{(0)}\over
1-F^{(0)} } \hat{H}_3 +
\end{equation}
$$
+ {1\over 2}\hat{E}^{++}\wedge \hat{E}^{--} \wedge
\left(e^{{1\over 2}\hat{\Phi}}F^{(0)} ~\hat{R}_1 -
e^{-{1\over 2}\hat{\Phi}}
\sqrt{1-(F^{(0)})^2}~
d\hat{\Phi} \right) + (d{{\cal L}}^{D1}_2)_{g} + \propto \hat{E}^I \wedge 
\hat{E}^J,
$$
where $d\hat{{\cal L}}_2^{IIB}$ is the external derivative of the
Lagrangian 2--form of the fundamental superstring
\p{dLIIBc} and $(d{{\cal L}}^{D1}_2)_{g}$ denotes the terms

\begin{equation}\plabel{dL2D1g}
(d{{\cal L}}^{D1}_2)_{g} = \left( dQ_0 +  e^{{1\over 2}\hat{\Phi}}
\hat{R}_1 -{1\over 2}Q_0d\hat{\Phi} \right)
\wedge
\left[ e^{-{1\over 2}\Phi}
\left(dA -\hat{B}_2\right) -{1\over 2}
\hat{E}^{++}\wedge \hat{E}^{--} F^{(0)}\right] +
\end{equation}
$$
-
\left( Q_0 +
{e^{-{1\over 2}\hat{\Phi}}F^{(0)} \over
\sqrt{1-(F^{(0)})^2}}\right)
\Big[{1\over 2}\hat{E}^{++}\wedge \hat{E}^{--} \wedge
dF^{(0)} + F^{(0)} e^{-{1\over 2}\hat{\Phi}} d{{\cal L}}^{IIB}_2
+
(1+F^{(0)})e^{-{1\over 2}\hat{\Phi}}\hat{H}_3\Big].
$$
which determine the dynamics of the gauge (super)field.

For the analysis of fermionic gauge symmetries and derivation of the
fermionic equations \p{dTh1D1c}, \p{dTh2D1c}
it is sufficient to consider the derivative of the
Lagrangian form $d{{\cal L}}^{D1}_2\vert_s$ taken on the surface
determined by the
superembedding condition \p{EID1c}, the gauge field constraint
\p{F=F} and the
algebraic equation  \p{Q0}.
(This means, in particular, that we can omit last two terms in
Eq. \p{dL2D1}).
After some straightforward but tedious algebraic manipulations with the
use of supergravity constraints \p{IIBc}, \p{H3c}, \p{R1c}, \p{R3c} one
obtains
\begin{equation}\plabel{2dL2D1}
e^{{1\over 2}\Phi} d{{\cal L}}^{D1}_2\vert_s =
\end{equation} 
$$
+i
\sqrt{{1+F^{(0)} \over 1- F^{(0)}}}
\hat{E}^{++}
\wedge
\left( \hat{E}^{2-}_{~\dot{q}} +
{\sqrt{1-F^{(0)} \over 1+ F^{(0)}}} \hat{E}^{1-}_{~\dot{q}}\right)
\wedge \left( \hat{E}^{2-}_{~\dot{q}} +
{\sqrt{1-F^{(0)} \over 1+ F^{(0)}}} \hat{E}^{1-}_{~\dot{q}}\right) +
$$
$$
- i
\sqrt{{1-F^{(0)} \over 1+ F^{(0)}}}
\hat{E}^{--}
\wedge \left( \hat{E}^{2+}_{~{q}} -
{\sqrt{1+F^{(0)} \over 1 - F^{(0)}}} \hat{E}^{1+}_{~{q}}\right)
\wedge \left( \hat{E}^{2+}_{~{q}} -
{\sqrt{1+F^{(0)} \over 1 - F^{(0)}}} \hat{E}^{1+}_{~{q}}\right) +
$$
$$
- {1 \over 2} \hat{E}^{++} \wedge \hat{E}^{--}
\wedge \left( \hat{E}^{2-}_{~\dot{q}} +
{\sqrt{1-F^{(0)} \over 1+ F^{(0)}}} \hat{E}^{1-}_{~\dot{q}}\right) ~
V^{+\underline{\a}}_{\dot{q}}
\left[
(F+{1\over 2} )\nabla_{\underline{\a}1}\Phi -
(F-{1\over 2} ){\sqrt{1+F^{(0)} \over 1- F^{(0)}}}
\nabla_{\underline{\a}2}\Phi
 \right]
$$
$$
- {1 \over 2} \hat{E}^{++} \wedge \hat{E}^{--}
\wedge
\left( \hat{E}^{2+}_{~{q}} -
{\sqrt{1+F^{(0)} \over 1 - F^{(0)}}} \hat{E}^{1+}_{~{q}}\right)
~V^{-\underline{\a}}_{{q}}
\left[ (F-{1\over 2} )\nabla_{\underline{\a}1}\Phi + 
(F+{1\over 2} ){\sqrt{1-F^{(0)} \over 1+F^{(0)}}}
\nabla_{\underline{\a}2}\Phi
\right].
$$

Bosonic coordinate equation of motion for super-D1-brane in general type
IIB supergravity background reads

\begin{equation}\plabel{MID1c}
\hat{M}^{I}_{2(D1)} \equiv
{1 \over \sqrt{1-(F^{(0)})^2 }}
 \hat{M}^{I}_{2(IIB)}  + 
\hat{E}^{++} \wedge \hat{E}^{--}
\left( U^{I\underline{a}} \nabla_{\underline{a}} \Phi -
U^{I\underline{a}} \hat{R}_{\underline{a}} e^{\hat{\Phi} }
\right) -
\end{equation}
$$
- {1 \over 2}  \hat{E}^{++} \wedge \hat{E}^{--}
\left[ \left( {\sqrt{1+F^{(0)} \over 1- F^{(0)}}}
e^{-{1\over 2} \hat{\Phi} } H_{\underline{a}\underline{b}\underline{c}}
+e^{{1\over 2} \hat{\Phi} } R_{\underline{a}\underline{b}\underline{c}}
\right) U^{++\underline{a}} U^{--\underline{b}} U^{I\underline{c}}
\right] -
$$
$$
- 4i \gamma^I_{q\dot{q}}
\left({\sqrt{1+F^{(0)} \over 1- F^{(0)}}}
\hat{ E}^{+q1} \wedge \hat{E}^{-\dot{q}1} -
{\sqrt{1+F^{(0)} \over 1- F^{(0)}}}
\hat{E}^{+2}_{q} \wedge \hat{E}^{-2}_{\dot{q}}
+ \hat{ E}^{+q1} \wedge \hat{E}^{-\dot{q}2} +
\hat{ E}^{+q2} \wedge \hat{E}^{-\dot{q}1}
\right)
$$
$$
+ \hat{E}^{++} \wedge
\left[
\hat{E}^{-1}_{\dot{q}}
\left( \nabla_{\underline{\a}2} \Phi +{\sqrt{1+F^{(0)} \over 1- F^{(0)}}}
\nabla_{\underline{\a}1} \Phi
\right) +
\hat{E}^{-2}_{\dot{q}}
\left(\nabla_{\underline{\a}1} \Phi -{\sqrt{1+F^{(0)} \over 1- F^{(0)}}}
\nabla_{\underline{\a}2} \Phi
\right)
\right] {\g}^I_{q\dot{q}}
V^{-\underline{\a}}_q
+
$$
$$
+ \hat{E}^{--} \wedge
\left[
\hat{E}^{+1}_{{q}}
\left( \nabla_{\underline{\a}2} \Phi +{\sqrt{1+F^{(0)} \over 1- F^{(0)}}}
\nabla_{\underline{\a}1} \Phi
\right) +
\hat{E}^{+2}_{{q}}
\left(\nabla_{\underline{\a}1} \Phi -{\sqrt{1+F^{(0)} \over 1- F^{(0)}}}
\nabla_{\underline{\a}2} \Phi
\right)
\right] {\g}^I_{q\dot{q}}
V^{+\underline{\a}}_{\dot{q}}=0 .
$$

\end{document}